# Walking Through Twitter: Sampling a Language-Based Follow Network of Influential Twitter Accounts

Felix Victor Münch, Leibniz Institute for Media Research | Hans-Bredow-Institut (HBI), Germany

Ben Thies, Humboldt University Berlin

Cornelius Puschmann, Centre for Media, Communication and Information Research (ZeMKI), University of Bremen, Germany

Axel Bruns, Digital Media Research Centre (DMRC), Queensland University of Technology (QUT), Australia

## Corresponding author

Felix Victor Münch; Leibniz Institute for Media Research | Hans-Bredow-Institut, Rothenbaumchaussee 36, 20148 Hamburg, Germany; f.muench@leibniz-hbi.de

## Abstract

Twitter continuously tightens the access to its data via the publicly accessible, cost-free standard APIs. This especially applies to the follow network. In light of this, we successfully modified a network sampling method to work efficiently with the Twitter standard API in order to retrieve the most central and influential accounts of a language-based Twitter follow network: the German Twittersphere. We provide evidence that the method is able to approximate a set of the top 1 to 10 percent of influential accounts in the German Twittersphere in terms of activity, follower numbers, coverage and reach. Furthermore, we demonstrate the usefulness of these data by presenting the first overview of topical communities within the German Twittersphere and their network structure. The presented data mining method opens up further avenues of enquiry, such as the collection and comparison of language-based Twitterspheres other than the German one, its further development for the collection of follow networks around certain topics or accounts of interest, and its application to other online social networks and platforms in conjunction with concepts such as agenda setting and opinion leadership.

## Keywords

Public Sphere, Social Media, Subgraph Sampling, Data Mining, Network Analysis, Influence, German Twittersphere, Follow Network, Twitter

# Introduction

Twitter is used by individuals, grassroots movements, and political and social elites to directly communicate to the public and influence opinion (see, e.g., Rogers, 2013). The platform appears relatively accessible to researchers because the majority of accounts post publicly and its Application Programming Interface (API) remains rather open in comparison with other Online Social Networks (OSNs). However, after Twitter started introducing increasingly restrictive rate limits and enforcing stricter terms of service regarding the sharing of data in 2012 (see, e.g., Puschmann & Burgess, 2014), the global follow network stopped being accessible to the majority of researchers (for exceptions, see, e.g., Myers et al., 2014).

This development resulted in a lack of independent research on a key mechanism for information diffusion and a global infrastructure for influence. After all, despite sponsored content and algorithmic sampling, follow or subscription networks are for many OSNs still a main predictor for content exposure. While it is widespread research practice to address this lack by using proxies for networks of attention on Twitter, such as mention, co-hashtag, or retweet networks (e.g., Himelboim et al., 2017), all of these rely on active communication. Therefore, widespread practices such as silent listening may be underrepresented.

This situation is aggravated by a recent change in how Twitter assigns account identification numbers (IDs), which allow the easy retrieval of information from its API. Its consecutive numbering scheme allowed a few independent, technically and monetarily costly projects such as the Australian Tracking Infrastructure for Social Media Analysis (TrISMA) (Bruns et al., 2016) to collect details of public accounts globally. Based on this, national follow networks could be captured and analysed (Bruns et al., 2017; Bruns & Enli, 2018). However, Twitter closed this possibility by assigning account IDs at random, undermining further data mining efforts following this collection strategy.

These technical restrictions place considerable limitations on research that aims to assess Twitter's relevance for issues such as news framing, opinion leadership and intermedia agenda setting processes (Iyengar et al., 2010; Scheufele & Tewksbury, 2007; Watts & Dodds, 2007). For example, Barberá (2015) derives a measure of individual ideological scaling based on the network position of Twitter users, demonstrating that the latter reliably predicts the former. Conway et al. (2015) are able to show that intermedia agenda setting takes place in U.S. politics, with the Twitter messages of political candidates in 2012 both predicting and echoing mainstream media messages, while Colleoni et al. (2014) investigate political homophily in U.S. politics in networks of reciprocated and non-reciprocated ties among the followers of the two main parties. In contradiction to widespread beliefs regarding online echo chambers as largely self-contained and insular environments, Valenzuela et al. (2017) find a greater effect of Twitter on television news than vice versa. However, all such studies have in common that in those cases where they describe processes *within* the platform they should be informed by a realistic view of the overall network.

Therefore, we present a large-scale test of a new Twitter follow network mining technique, building on the so-called rank degree method (Salamanos et al., 2017a, 2017b, 2017c; Voudigari et al., 2016), which we describe in the Methods and Analysis section below. As this walk-based technique only requires local information to sample a graph, we were able to adapt it as a data mining method for the follow networks of influential Twitter users, using the cost-free Twitter standard APIs. This approach has been tested by using the method to draw a sample of the German-speaking Twittersphere.

Our analysis shows that the sample exhibits influence and activity measures orders of magnitude higher than a random sample of the same size from a near-complete collection of German-using Twitter accounts from the 2016 TrISMA dataset (Bruns et al., 2016). This is evidence that our sample represents an influential backbone, an approximation of the proverbial, highly influential 'top 10 percent' of this language-based Twittersphere. A test study employing community detection and keyword extraction shows that this network sample is suitable for investigating large-scale topical communication structures, corresponding with issue publics or communities in a language-based Twittersphere.

In light of the continuously tightening restrictions on the Twitter standard APIs, our adaptation of the rank degree method for gathering network samples is a valuable alternative to more brute-force approaches, as it can be implemented by small teams at low costs in terms of time and budget. Even though Twitter has made it almost impossible for independent researchers to capture comprehensive, large-scale follow networks, our method works around these restrictions to produce an overview of the overall structure of such networks, in this case, for the German Twittersphere.

## *Opportunities Opened up by Follow Network Samples*

Nation-level data about follow networks enable a multitude of avenues of enquiry. Amongst others, they make further data collection possible, such as constant monitoring of the most influential accounts, without being restricted to single topics. For example, this allows for a platform-independent assessment of trending topics and public opinion on a platform, of behavioural changes that are the result of adjustments made to recommendation algorithms, or of the positions, roles, and influence of automated accounts.

Combined, these data support the further development of media and communication theory as well as social theory regarding networked public spheres on an empirical basis. In the case of Australia, this has already been shown by Bruns & Highfield (2016): they can base their reappraisal of public sphere theory by Habermas (1962) – in the form of a more up-to-date concept of a networked public sphere – on a complete collection of the Australian Twitter follow network, detectable topical communities, and the localised spread of hashtags within it.

Also, research methods and paradigms that have their roots in more qualitative practices can benefit from this kind of large-scale data. For example, Dehghan (2018) grounds his discourse analysis of polarising discussions on Twitter about the Australian Racial Discrimination Act on the same dataset.

Beyond academic research, the usefulness of large scale social media data of this kind in a political or commercial context is clearly given. In public relations and (influencer) marketing, the benefit of being able to get an overview of the communication landscape of an OSN is obvious. Furthermore, finding answers to questions of (media) policy, for example, regarding the fragmentation of the media audience as described by McQuail (2010, pp. 444–445), can be supported with these data.

The hypothesised fragmentation of online news audiences represents another area of research in which a holistic view of national Twitterspheres can yield decisive results, in some cases allowing inferences about news consumption that extend beyond social media. For example, there is evidence for shared patterns of public attention that reach from social media to mass media on a transnational level and exhibit fragmentation that is accompanied by a high degree of audience duplication in select contexts (Fletcher & Nielsen, 2017; Webster & Ksiazek, 2012; Tewksbury, 2005). Such research can be augmented by relating the news sharing behaviour on Twitter to the network structure to determine whether online news audiences are fragmented along network community structures or, conversely, no visible relationship between news preferences and network structure exists.

## *Problem: Restricted API Access for Researchers to Gather Follow Network Samples*

This kind of data, however, has become less and less accessible in the past years. In part this has resulted from increasingly strict privacy legislation (such as the European GDPR) taking effect and the subsequent arrangements that Twitter has made to comply with such regulation. On the other hand, the primary purpose of Twitter's APIs has never been to support academic research, but to enable developers to build products that make use of Twitter data. Therefore, information that is not necessarily needed for such products may be withheld or not even stored, rather than being made accessible to the research community.

In fact, Twitter offers three different kinds of APIs – the standard, the premium, and the enterprise APIs (Tornes, 2017) – and the free standard APIs are the most restrictive. The Standard APIs allow its users to retrieve account information and also to query tweets. For both functions API calls are limited and have a cooldown time. For example, when looking up the friends[1] of a user, a single API user may execute a maximum of 15 calls every 15 minutes, retrieving 5,000 friends each call – so a maximum of 75,000 friends of 1 to 15 accounts can be retrieved within 15 minutes.

The main advantages of the premium and enterprise APIs are higher rate limits for receiving tweets as well as access to a longer history of tweets and account activity-related features[2]. However, researchers face affordability and accessibility issues with both of these services, as access to the enterprise API is only sold through direct contact with a salesperson (prices are not mentioned online), and access to the premium API is also subject to case-by-case approval. The latter seems to be more restrictive than for the Twitter Standard API.

In the past, researchers were able to make use of Twitter's cost-free standard API to gather large collections of Twitter accounts, culminating in a collection by Bruns et al. (2016). Their collection method exploited the fact that Twitter assigned consecutive account IDs. It queried the free API for every consecutive possible account ID, gathering data for almost every global account in 2016. Twitter has since changed its policy of assigning account IDs to a random system. This involves much higher numbers than there are Twitter accounts, and thus renders more recent collections using this method impossible.

The APIs are also undergoing continuous changes, as certain user account properties are becoming protected, thus not accessible to researchers via the API anymore. In the past, these properties included geotags, user time zone, and the interface language used on Twitter, which are all inaccessible by now.

## *Objectives: Test of a Sampling Method and Data Mining of the German[3] Twittersphere*

Given those restrictions, this project's purpose was two-fold: first, we were testing an adaptation of a sampling method for influential nodes in a network that has shown promising 'lab' results (see below) as a data mining method 'in the wild' – using the cost-free Twitter standard APIs only. Our focus here was especially to test the practical feasibility of the method for a small research team with limited resources and to explore which adaptations have to be made for it to work under this objective. Furthermore, we also probed the usefulness of the so-gained sample for some of the research opportunities mentioned above, by identifying topical communities amongst most influential accounts in the German-speaking Twittersphere.

Second, this project's objective was to open up further avenues of enquiry by either providing data on the German Twittersphere for other projects, or, (especially when this is not possible while staying compliant with Twitter's terms of service) to provide other researchers with the method and its implementation, i.e. the code for a prototype, for sampling either the same data or other language-based Twitterspheres. It can be found under an open source license in the supplementary materials of this paper as well as in an online git repository[4]. We invite other interested parties in supporting us with its further development there.

---

[1] Here and in the following, we refer to a user that follows another user on Twitter as follower and to the user being followed by another user as a friend (so in this case, friendship is not necessarily mutual).
[2] https://developer.twitter.com/en/products/products-overview, retrieved on 5.7.2019
[3] For this article, except otherwise stated, the term German stands for the language and not a national affiliation.
[4] https://github.com/FlxVctr/RADICES

# Background

While Twitter does not represent the general population of a country or language domain, it is an interesting population in itself. This has motivated a number of successful attempts to sample or completely collect other national Twitterspheres. However, all these attempts were either not based on the Twitter follow network or relied on properties of the Twitter standard APIs that are no longer available.

## *Representativeness of Twitter Data and Representativity of Social Network Samples in General*

A common criticism of Twitter research, and social media research in general, is that its users are not representative of the general population of a country or language domain at large (e.g., Blank, 2017; Mellon & Prosser, 2017). We do not claim this kind of representativity. However, Twitter users by themselves represent a population of interest in many countries worldwide. Even in Germany, where Twitter plays a comparatively smaller role (it is ranked 16th in Germany according to Alexa[5], compared to rank 8 in the USA[6] and rank 11 worldwide[7]), 4 percent of the German-speaking population over 14 are using Twitter at least once per week and this value appears to be stable over the last few years according to representative surveys (Frees & Koch, 2018). This makes Twitter a niche network, compared to competitors like Facebook and Instagram. However, its reach is comparable with the 3 percent of weekly users of audio podcasts in Germany in 2018 (Frees & Koch, 2018), or 4.5 percent who had a subscription to a national newspaper in Germany in 2017 (Pasquay, 2018). Moreover, the results of a study by Nordheim, Boczek, & Koppers (2018) of broadsheet newspapers in the United Kingdom (The Guardian), the United States of America (New York Times), and Germany (Süddeutsche Zeitung), indicate not only that social media “resurged massively” (von Nordheim et al., 2018) as a news source between 2009 and 2016, but also that Twitter is more commonly used as a news source by journalists than Facebook and represents an “elite channel” (von Nordheim et al., 2018). In accordance with this assessment, Wojcik & Hughes (2019) found that in the USA highly educated, wealthy people under 49 are over-represented on Twitter, while Blank (2017) makes similar observations regarding the UK. While Twitter users in the U.S. were more Democratic leaning than the average US adult, race and gender distribution showed only minor disparities with the population average, while women and users who tweet about politics were overrepresented within the most active 10 percent of Twitter users.

These top 10 percent being responsible for 80 percent of the tweets within the dataset analysed by Wojcik & Hughes (2019) points to another issue with representativity in OSNs. It lies in the fact that OSNs usually exhibit heavily skewed distributions regarding activity and connectivity. This means that a traditional understanding of representativity based on common statistical approaches, which often assume normal distributions, will not be useful. Consequently, in this paper we move our focus in terms of sample quality from typicality (i.e. traditional representativeness) to another goal: getting the most influential accounts and a backbone structure of the network at the smallest possible cost.

[5] https://www.alexa.com/topsites/countries/DE, retrieved on 27.6.2019
[6] https://www.alexa.com/topsites/countries/US, retrieved on 27.6.2019
[7] https://www.alexa.com/topsites, retrieved on 27.6.2019

### *Related Research: Location- or Language-Based Twittersphere Collections*

This project is preceded by and builds on previous successful endeavours to collect other language- or nation-based Twitterspheres. To our knowledge, the first project mapping a national Twittersphere was conducted by Ausserhofer & Maireder (2013). However, while innovative at the time in their methods and with impressive results given the exploratory nature of their study, they analysed @-mention networks of only a few hundred accounts whose collection was based on keywords related to Austria. Geenen, Boeschoten, Hekman, Bakker, & Moons (2016) followed a similar approach for the Netherlands, also analysing @-mention networks based on a keyword search for specifically Dutch terms. Ausserhofer & Maireder (2013) argue that @-mentions would be a ‘better’ measure of influence than follower numbers. However, this claim may be an overinterpretation from a study by Cha, Haddadi, Benevenuto, & Gummadi (2010), which showed that for the top 10 percent of Twitter accounts in terms of follower numbers the number of retweets and replies was only weakly rank-correlated with the follower numbers. Closer inspection of this study also shows that retweet- and mention-based influence fluctuates over time and by topic. In any case, while @-mention networks are easier to collect, because their collection is less restricted by the Twitter standard APIs, the explanatory power if used alone depends on the definition of influence and neglects the influence of highly followed accounts on silent listeners on Twitter. Especially because a reaction to a tweet with an @-mention or retweet depends on having seen this tweet first, they should be analysed in combination with follow networks.

Bruns et al. (2017) presented an analysis of a follow network based on a comprehensive collection of Australian Twitter accounts in 2016 by TrISMA (Bruns et al., 2016). As explained above, this collection was possible due to the fact that Twitter assigned user IDs in a consecutive way. The global dataset could then be filtered down to accounts who are likely to be located in Australia, and all their connections could be collected, even though with great effort. While the same method allowed the analysis of the Norwegian Twittersphere (Bruns & Enli, 2018), this method has been rendered impossible by now.

## Methods and Analysis

In this section we will first describe our sampling method, which we based on the so called rank degree method (Voudigari et al., 2016), adapted for directed networks, and optimised for practicality and efficiency with the cost-free Twitter standard APIs. After this, we describe our assessment of the sample’s quality in accordance with our sampling goals: the collected accounts’ influence as measured by coverage, reach, activity, and follower numbers. Finally, we present methods and results of a test study using our sample, which yields promising results: by means of community detection, keyword extraction from tweets, and a manual inspection of account descriptions, we are able to detect topical communities within this highly influential sample of the German Twittersphere. Our results showcase an intuitive overview of the structure of this part of the German networked public sphere, that resembles preceding analyses of the Australian Twittersphere.

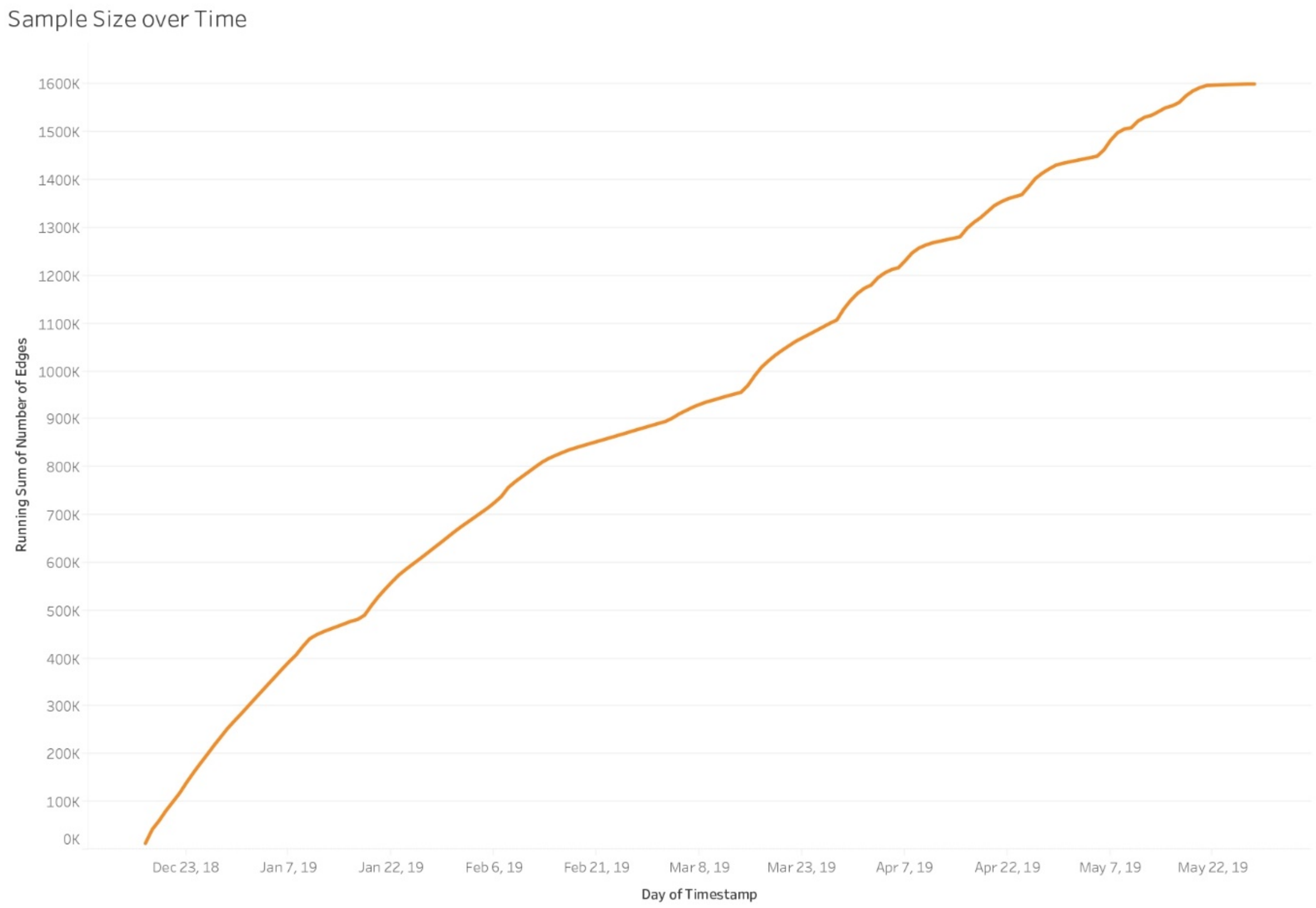

***Figure 1:*** *Sample size as measured by the total number of edges over time (18-12-17 – 19-05-28)*

## Sampling

Our sample was collected from the mid of December 2018 until the end of May 2019 (Figure 1). While we have almost reached our initial goal of collecting 1 million accounts (937,809 accounts were collected; the goal was mostly determined by the time and resources available for the overall project and represents about 5 percent of the number of German using accounts as determined by TrISMA), our collection was stopped by the fact that the Twitter standard API was changed and the interface language of an account, on which our method relied, was made a private property. However, our method can and will be changed so that it infers a user language by identifying the Tweet language(s).

*The Rank Degree method***.** To draw a sample of the most influential accounts within the German Twittersphere, we chose to develop and test a modification of the rank degree method.

The rank degree method (Voudigari et al., 2016) is a mostly deterministic, walk-based algorithm that only requires local information to sample a graph. Assume an undirected network (or a directed network in which all edges are reciprocal), given initial nodes as seeds and a desired sample size *x*. Then the rank degree process works as follows:

1. For each seed *w*, find the connected node *v* with the highest degree.
2. Update the sample with the selected edges (*w, v*) and the symmetric one (*v*, *w*).
3. Update the source graph by removing the edges selected in 2.
4. Update seeds so that only the new node *v* for every *w* is kept.

5. If all *v*s in the seeds are leaves (degree = 1), select new initial seeds.
6. Repeat 1. – 7. until sample is of size *x*.

From the steps outlined above, it becomes evident that the process is deterministic in that it completely depends on the initial seeds chosen (and, in case 5. is executed, on the new initial seeds). Moreover, any selected node can be visited multiple times, but each time from a different node. If any two walkers visit the same node at the same time, those two walkers will collapse and further proceed as one. By updating the graph every time, the process ultimately alters the nodes' degree and ranking, resulting in a dynamic sampling process.

Note that there exist two versions of this algorithm. The first version works as outlined above, while with the second version, instead of only the top connected node in step 1, the top *k* connected nodes would be selected, where *k* is calculated by $\rho$ * #connections($w$), $0 < \rho \leq 1$. The first version has been shown to produce better samples in general, and Salamanos et al. (2017a, 2017c) discuss the algorithm in this form. Comparing the algorithm to other graph sampling methods such as Forest Fire, Frontier Sampling, Metropolis Hastings and random sampling methods, they find that samples generated by their algorithm preserve several graph properties to a large extent and that influential spreaders can be identified at almost the same accuracy as in the full information case when having explored only 20 percent of the network.

*Our adaptation and implementation of the Rank Degree method***.** With any k > 1 the algorithm would generate seeds in an exponential manner. This would render its execution with the cost-free Twitter standard APIs practically impossible. While the use of k = 1 is still within the original definition of the rank degree method, further modifications to the algorithm for the purpose of mining Twitter were necessary since the rank degree method was not defined for directed networks by Voudigari et al. (2016) or Salamanos et al. (2017a, 2017b, 2017c). Moreover, as Coscia & Rossi (2019) point out, in real-world data collections, the assessment of the quality and performance of an algorithm has to balance efficiency regarding the API restrictions with sample quality. Accordingly, we made a number of adjustments, which we explain in this section, that mainly serve a more efficient collection and easier parallelisation in possible future developments.

In contrast to Voudigari et al. (2016) and Salamanos et al. (2017a, 2017b, 2017c), the full network is not available to us, therefore we had to access the required information directly from the Twitter standard API. While it was possible to increase the collection speed by using twelve API keys provided by personal accounts of the authors and other project contributors at our institutions, all API endpoints have a quota on a calls per time-window and per API key basis.

Before starting the sampling algorithm, we had to define a *seed pool*, a collection of Twitter IDs that can be randomly sampled from to use them as initial seeds (cf. *The Rank Degree Method*). In our case, we used a list of all approximately 15 million Twitter accounts who had their interface language set to German according to the dataset off all Twitter accounts in 2016 as collected by TrISMA (Bruns et al., 2016). The mining process was then set in place as follows:

1. From the *seed pool*, draw a random account *w*.
2. Look up the last 5,000 friends of *w* and rank them by their follower count.
3. Choose the friend *v* with the highest follower count to whom there is an unburned edge and whose interface language is set to German.
4. If *w* has no friends or if all outgoing follow connections were already burned, jump to a randomly drawn seed from the seed pool
5. Update the sample with the selected follow connection (*w, v*) and the symmetric follow connection (*v*, *w*) *if it exists.*
6. 'Burn'/save the follow connection so that it cannot be walked again.

7. Repeat steps 2. - 7. with *v* as the new starting point.

To use the full capacity of the API calls available to us, we used 200 parallel walkers. Contrasting to the original rank degree method, we do not let the walkers collapse when they land on the same node for time efficiency reasons but let them execute consecutively. While close to the feasible maximum with the API restrictions in place, 200 is a massively lower number of walkers than what has been used in the tests of the original rank degree method, which was 1 percent of the total number of nodes. In our case this would equal over 25,000 walkers[8].

This is not the only adaptation that we decided to make for practicality reasons. Instead of choosing the friend with the highest degree, i.e. the one with the most connections, it (1) only looks up the last 5,000 friends and then (2) chooses the one with the most followers if (3) it has its interface language set to German.

Adaption (1) was done for three reasons. First, it takes exactly one API call to access up to 5,000 most recent friends of an account. By default, only 15 API calls every 15 minutes are allowed for this endpoint per API key. Since there exist Twitter accounts with 50,000 friends and more, the time needed would have been increased unjustifiably for those accounts. As every account can pay only limited attention to its friends, the more friends one account has, the less they count individually. Therefore, including all friends of an account that follows several thousand accounts, would give those connections an importance which they most likely do not possess. This renders spending precious API calls beyond a 5,000 friends limit unjustifiable from our perspective. Third, while the original rank degree method for undirected networks uses the degree, we use the follower number only as the (approximate) in-degree, because the number of followed accounts, or out-degree, is no indicator of an account's influence (Cha et al., 2010).

Adaption (2) is a simplicity trade-off, acknowledging the fact that our sampling method can only be an approximation of the rank degree method. We do not possess knowledge about the whole network, and the usage of the follower number is only a heuristic for assessing influence, an approximation of the in-degree in our network of interest, the German Twittersphere. The actual in-degree within the German network is, in some cases, actually lower, due to non-German followers, and subject to change during the months of our collection. Therefore, dynamically adapting the degree was neglected in favour of an easier parallelisation of the walkers and a cache database of fixed account details, such as the friend connections and follower numbers, that led to a significant speedup of our collection.

Lastly, since our aim was to collect a sample of the German Twittersphere, we used the accounts' interface language as a filter criterion, hence (3).

Another significant change to the original algorithm is made in step 4. Voudigari et al. (2016) and Salamanos et al. (2017a, 2017b, 2017c) only apply their method to undirected networks, so the symmetric connection is always added to the sample in their case. In the directed case, there is not necessarily a symmetric connection, but, as in the original Rank Degree method, we add it, if there is one. This is done because we expect this to ensure that accounts with only a few but high in-degree followers will receive more representative scores regarding centrality measures such as Page Rank, Betweenness or Eigenvector centrality. Note, however, that, as in the original rank degree method, even though both edges are added to the sample, one of them ((*v, w*)) can still be walked, as only the directed edge is burned. Figure 2 illustrates the sampling procedure.

[8]When taking the 4 percent of weekly active Twitter German-speaking users in Germany (as determined by Frees & Koch (2018)) as a (low) estimate for the total number of German accounts

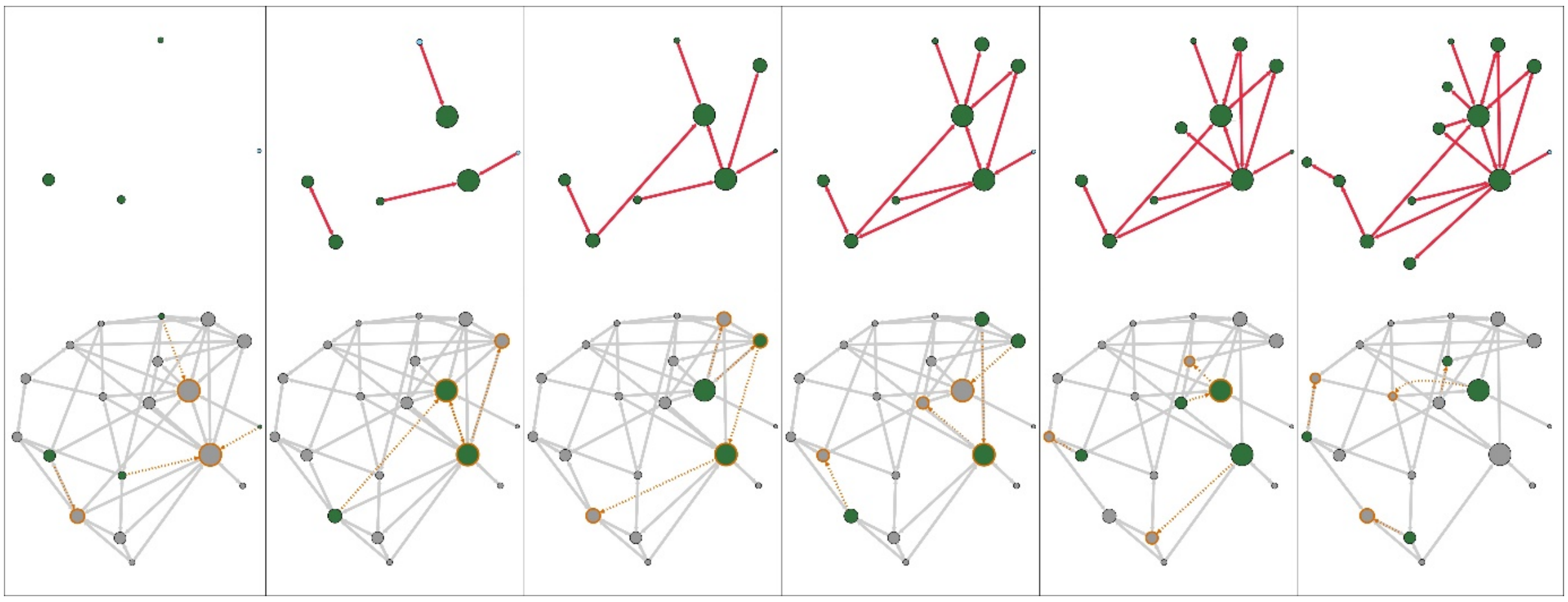

***Figure 2:*** *Our adaptation of the rank degree algorithm. The top panel represents the sample after every iteration, the bottom panel represents the underlying network without the removed edges. The example network is based on a student interaction network (Heidler et al., 2014), filtered for indegree > 3, as available from https://github.com/gephi/gephi/wiki/Datasets.*

## *Evaluating the Sample Quality*

As we do not possess knowledge about the whole network, assessing the sample quality, as done for the original rank degree method, was not possible for the German follow network. Neither could we expect or were aiming for any kind of typicality of properties of the sampled accounts in comparison to a population as would be necessary for a representative sample in a traditional sense. Our aim was to approximate the proverbial 1 to 10 percent at the top of the German Twitter population which are characterised by orders of magnitude higher activity, coverage, reach, and follower numbers than the remaining 99 to 90 percent.

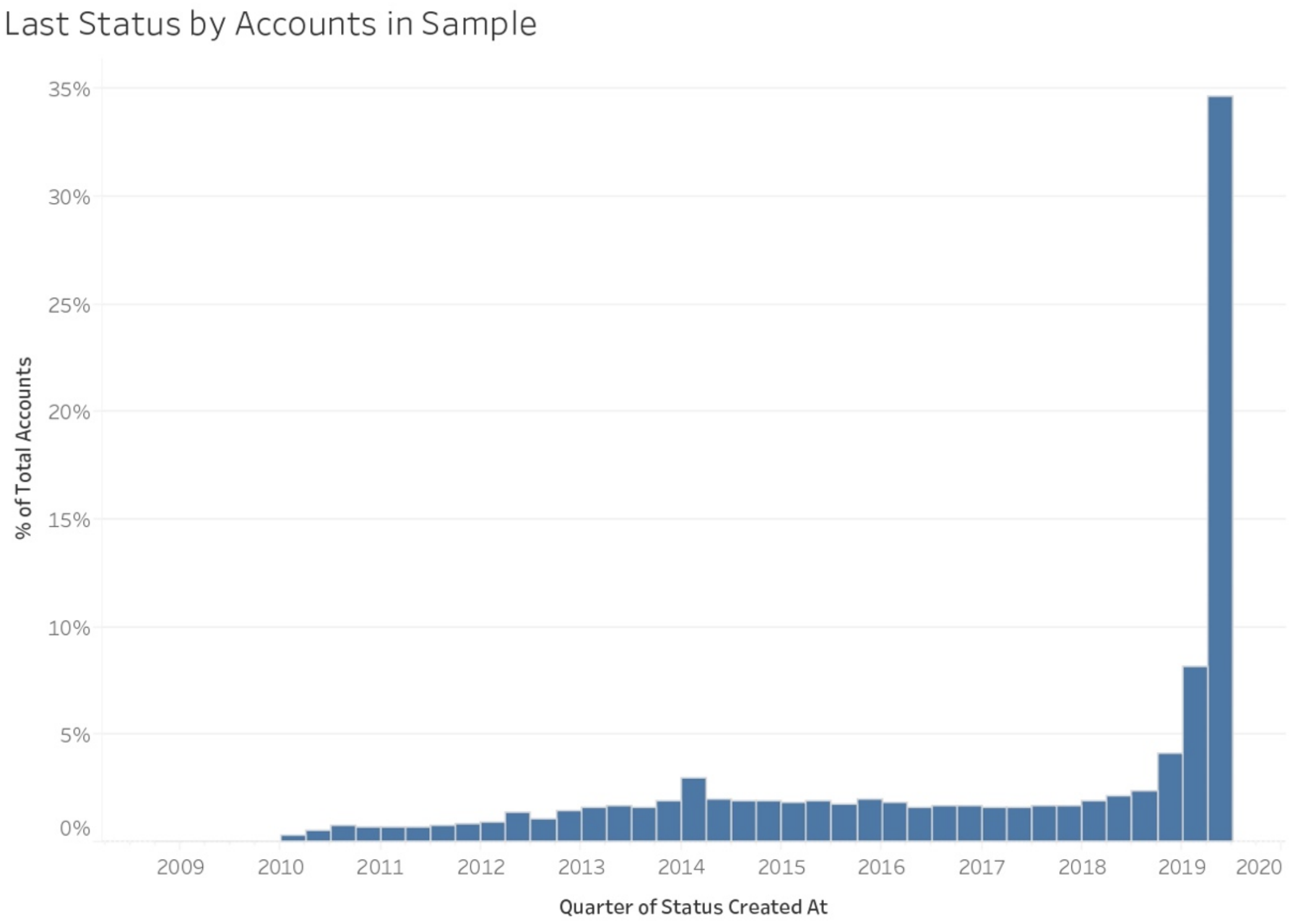

***Figure 3:*** *Distribution of the date of the last status by accounts in our sample at the end of the network collection timeframe (May 2019)*

*Activity and Centrality***.** As the seeds are just randomly selected in order to find influential accounts, we excluded those seeds from the sample that do not have an incoming edge from another node for assessing the activity and follower numbers of our sample. This filtering left us with about 197,000 unprotected accounts, whose activities and follower numbers we could access. As can be seen in Figure 3, even in this sub-sample, there is a large number of accounts who have not been active for years. Whether this is due to actual inactivity or simply silent usage of the platform cannot be determined here. However, over 42 percent of our sample have posted at least one tweet from the beginning of 2019 until the end of our network collection in May.

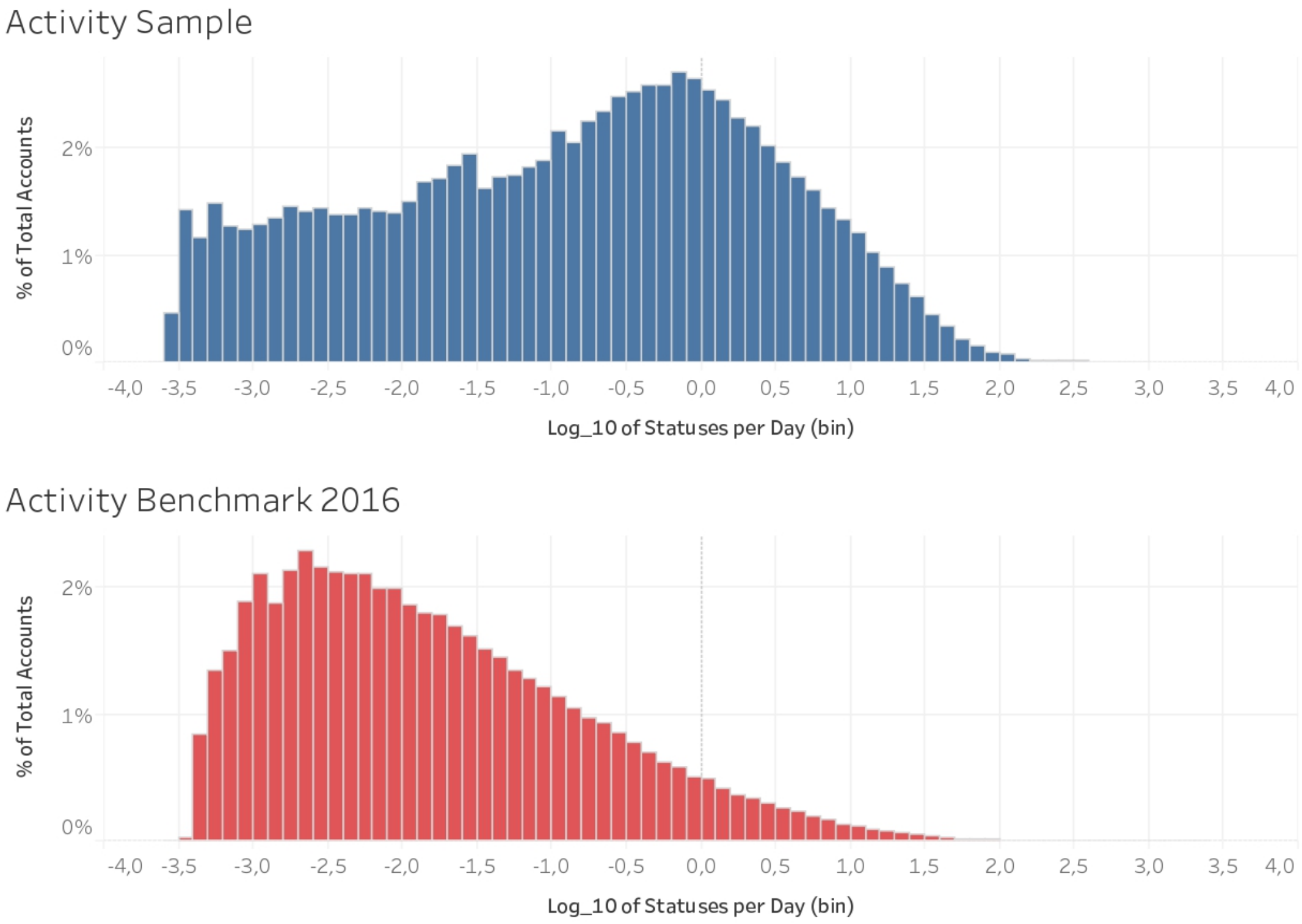

***Figure 4:*** *Comparison of our sample ('Sample') with all accounts collected in 2016 by TrISMA ('Benchmark 2016') regarding the distribution of the statuses per day since account creation.*

Nevertheless, as depicted in Figure 4, where we compare the distributions of the number of tweets per day since the account creation day by accounts in this sample with the same data in the subset of German-using accounts from the TrISMA collection, we can see that there is a pronounced qualitative difference in activity. The accounts in our sample are orders of magnitude more active in terms of tweets per day than this benchmark.

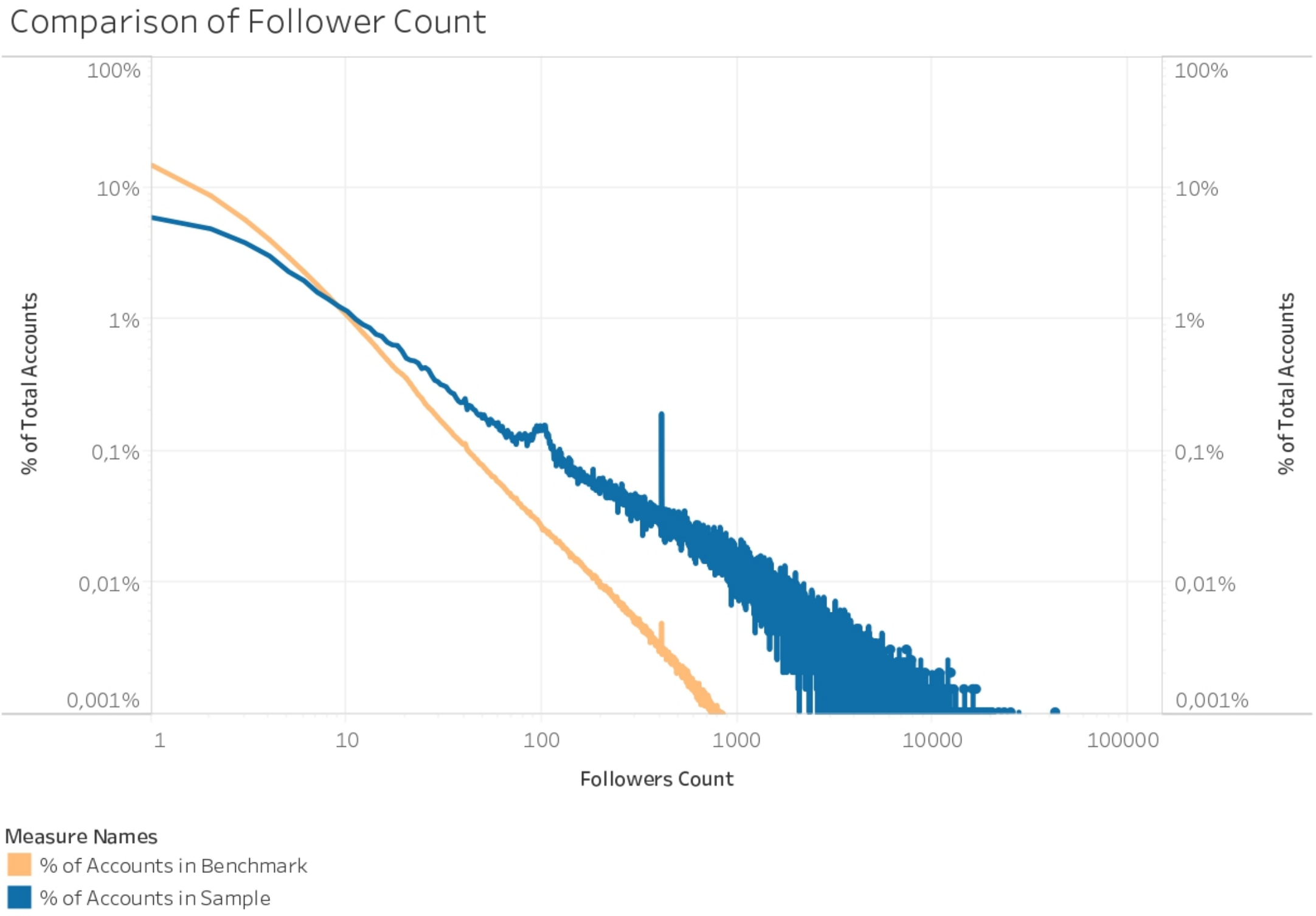

***Figure 5:*** *Comparison of our sample with all accounts collected in 2016 by TrISMA regarding the distribution of the follower count at the time of the sample collection. The spike between 100 and 1000 accounts is caused by a fully connected bot-net.*

A similar picture is drawn if we inspect follower numbers: Again, Figure 5 shows a comparison between our sample (without seeds that remained leaves) and the entirety of Twitter accounts in 2016 that had set their interface language to German. Here too, the distributions of follower numbers show a substantial qualitative difference, with the typical follower numbers in our sample being multiple times higher than in the benchmark.

In summary, this sample exhibits indeed a high-influence profile in terms of activity and in-degree centrality (as measured by the follower numbers reported by the Twitter API). We therefore will refer to it as the influencer sample from now on.

*Coverage and Reach.* However, activity alone does not translate to influence in terms of content exposure. Therefore, we tested what we call the 'coverage' of our influencer sample: the typical percentage of a German-using Twitter account's friends that are in our sample. Again and for the same reasons as above, we filtered out seeds that have no incoming edges in our sample. However, as we did not require protected information this time, the influencer sample size remained slightly higher at about 199,000 accounts for the following tests.

For this purpose, we drew a random sample of 1,000 accounts from the German TrISMA collection and retrieved their actual friends (including those with another language than German as interface language) from the Twitter API. From here on this sample will be called the test sample. Of course, the final size of this test sample was reduced due to deleted and protected accounts. Furthermore, we excluded accounts with less than 2 friends to avoid misleading coverage values of 100 percent and 0 percent.[9]

---

[9] We additionally did the same analysis including all accounts in the test sample. The results were comparable, with a mean of 36 percent and a median of 33 percent coverage.

| n = 597 | number of friends | percent of friends in influencer sample | percent of friends in baseline sample |
|---|---|---|---|
| **mean** | 57.1 | 40.2 | 0.5 |
| **std** | 160.4 | 30.3 | 2.7 |
| **min** | 2 | 0 | 0 |
| **25%** | 7 | 11.4 | 0 |
| **50%** | 18 | 40 | 0 |
| **75%** | 42 | 64.7 | 0 |
| **max** | 1988 | 100 | 50 |

***Table 1:*** *Count, mean, standard deviation, minimum, quartiles, and maximum of the number of friends and the percentages of friends in the influencer and baseline sample for public accounts in the test sample with at least 2 friends.*

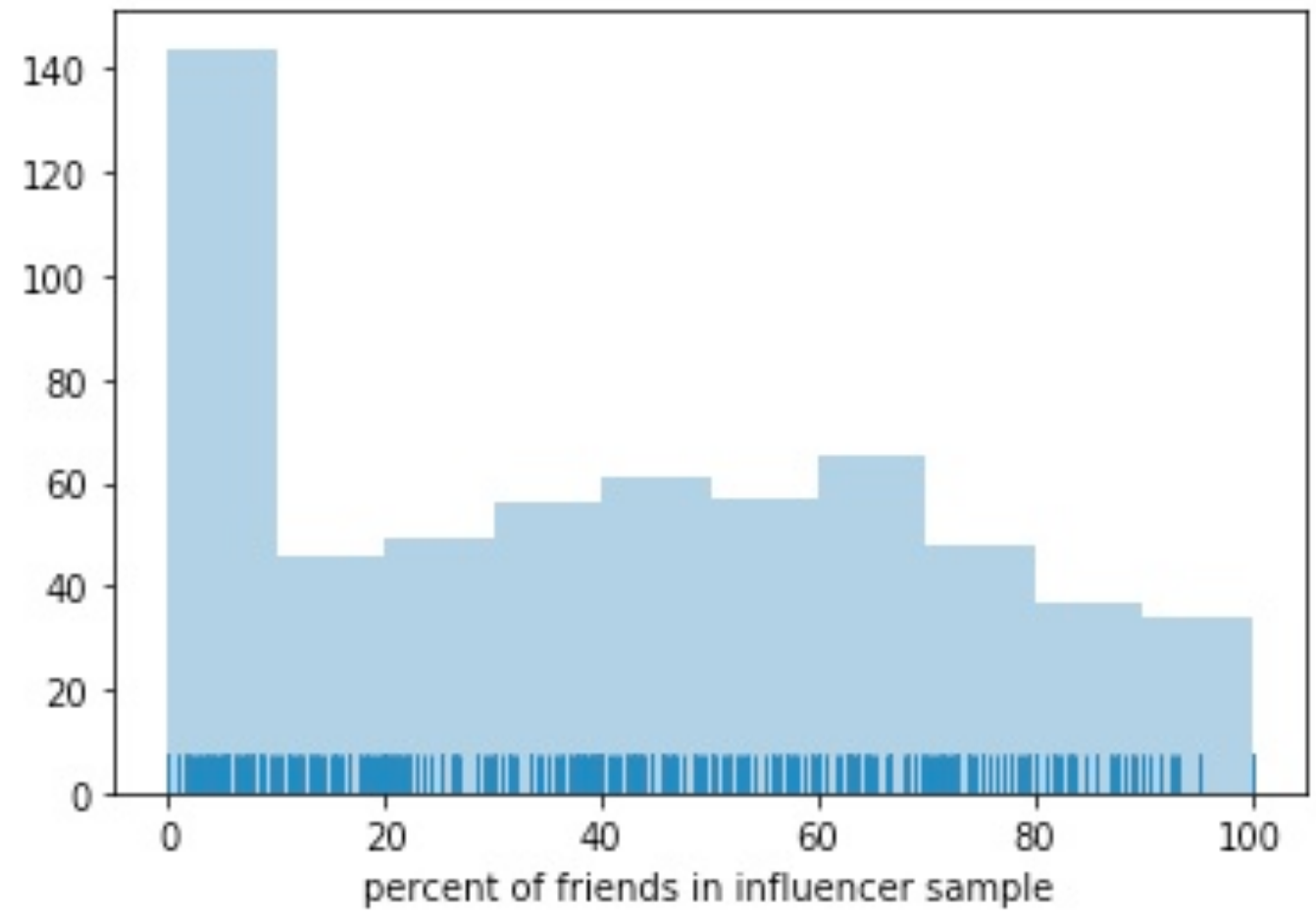

***Figure 6:*** *Distribution of accounts in the test sample over the percentage of their friends that can be found in the influencer sample (filtered for in-degree >= 1, leaving 199,180 accounts).*

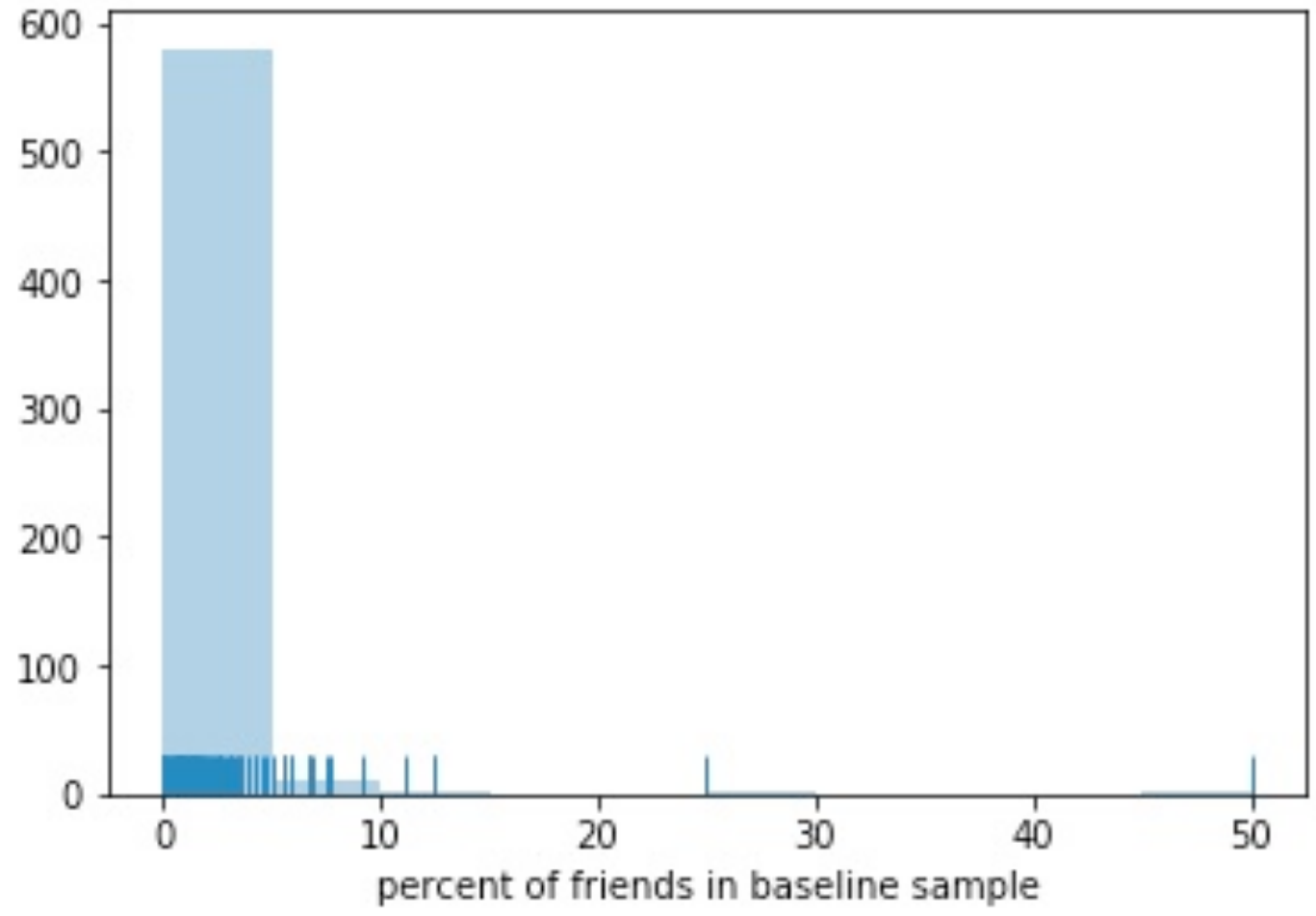

***Figure 7:*** *Distribution of accounts in the test sample over the percentage of their friends that can be found in the baseline sample (199,180 accounts drawn randomly from German-using accounts in TrISMA collection).*

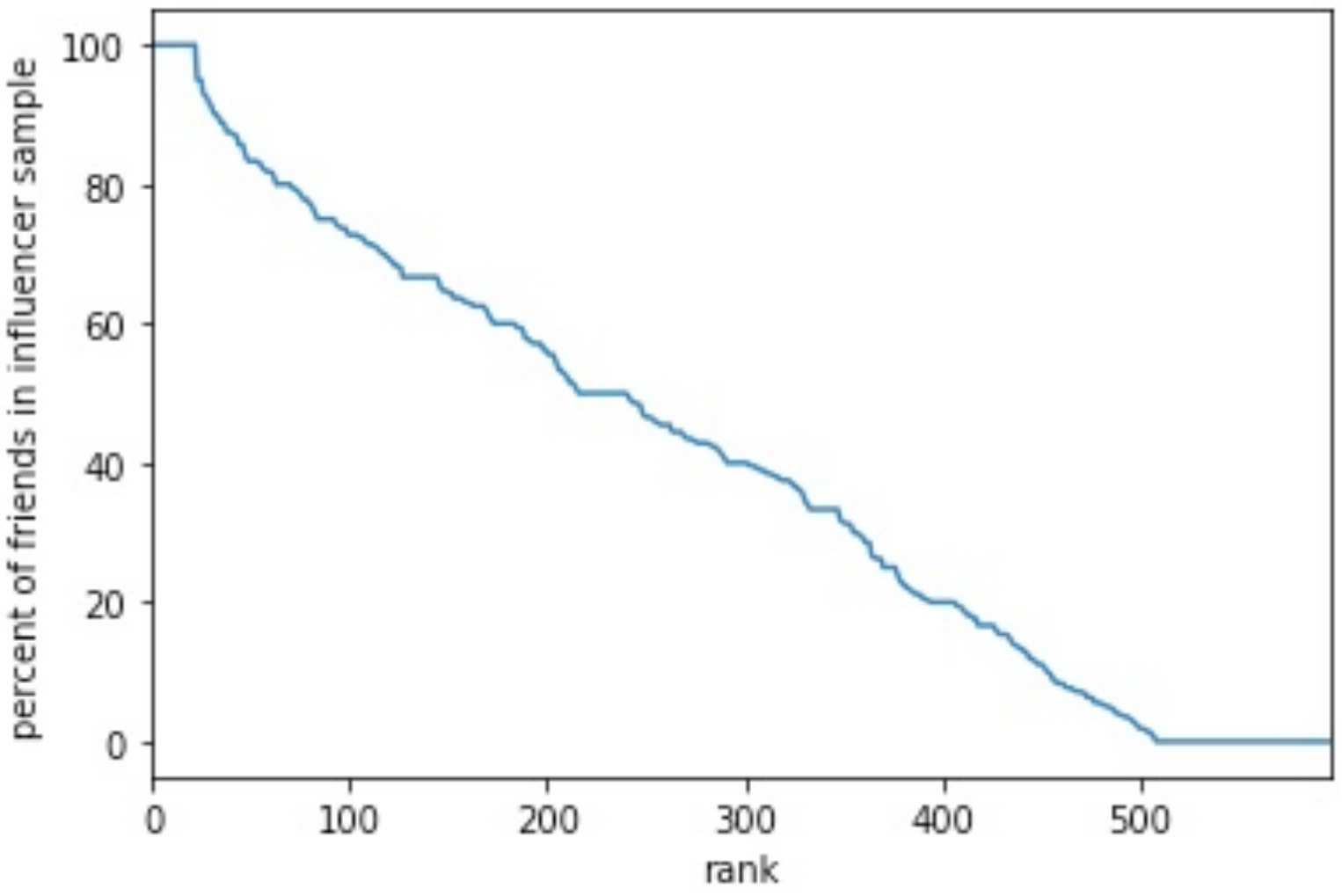

***Figure 8:*** *Rank-coverage distribution of accounts in the test sample with at least 2 friends for the influencer sample (filtered for in-degree >= 1, leaving 199,180 accounts)*

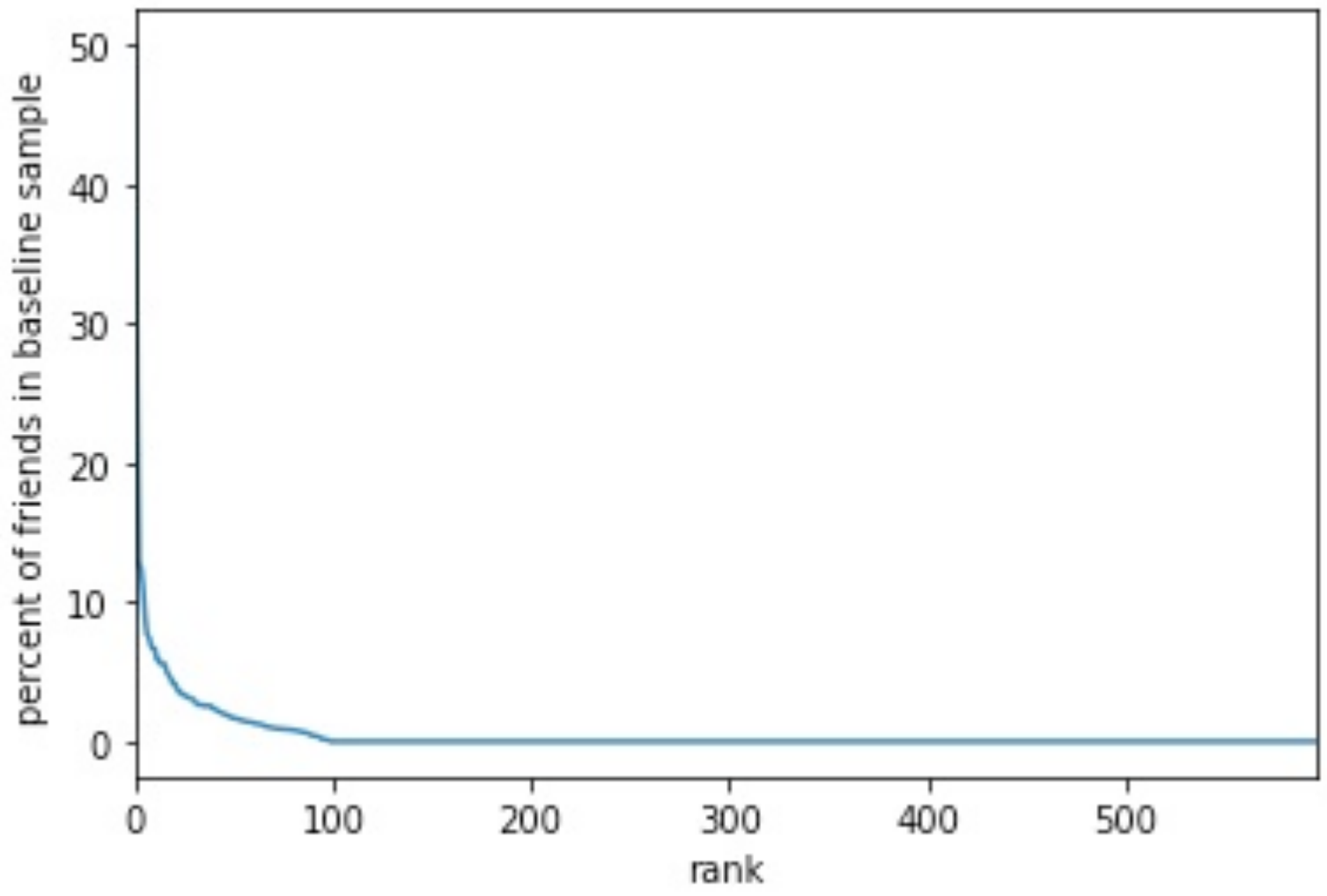

***Figure 9:*** *Rank-coverage distribution of accounts in the test sample with at least 2 friends for the baseline sample (199,180 accounts drawn randomly from German-using accounts in TrISMA collection)*

As a baseline we drew a random sample from the German-using accounts in the TrISMA collection with the size of our influencer sample (199,000 accounts). Then we evaluated the coverage of the influencer and the baseline sample for accounts in the test sample.

As can be seen in Table 1, the mean and median of the coverage of our influencer sample is at 40 percent, compared to 0.5 percent mean and almost 0 percent median coverage of the baseline sample. However, as distributions are often heavily skewed in networks, mean and median do not tell the whole story. As can be seen in the distribution plots in Figures 6-9, our influencer sample differs extremely from the baseline sample in terms of coverage distribution, so it is evident that we observe a different class of accounts here. When ignoring accounts with 0 percent coverage, for the influencer sample, Figure 6 shows a distribution of coverage resembling a normal distribution around the mean/median of 40 percent. In other words, on average, 4 out of 10 friends of a German-using Twitter account are in our sample.

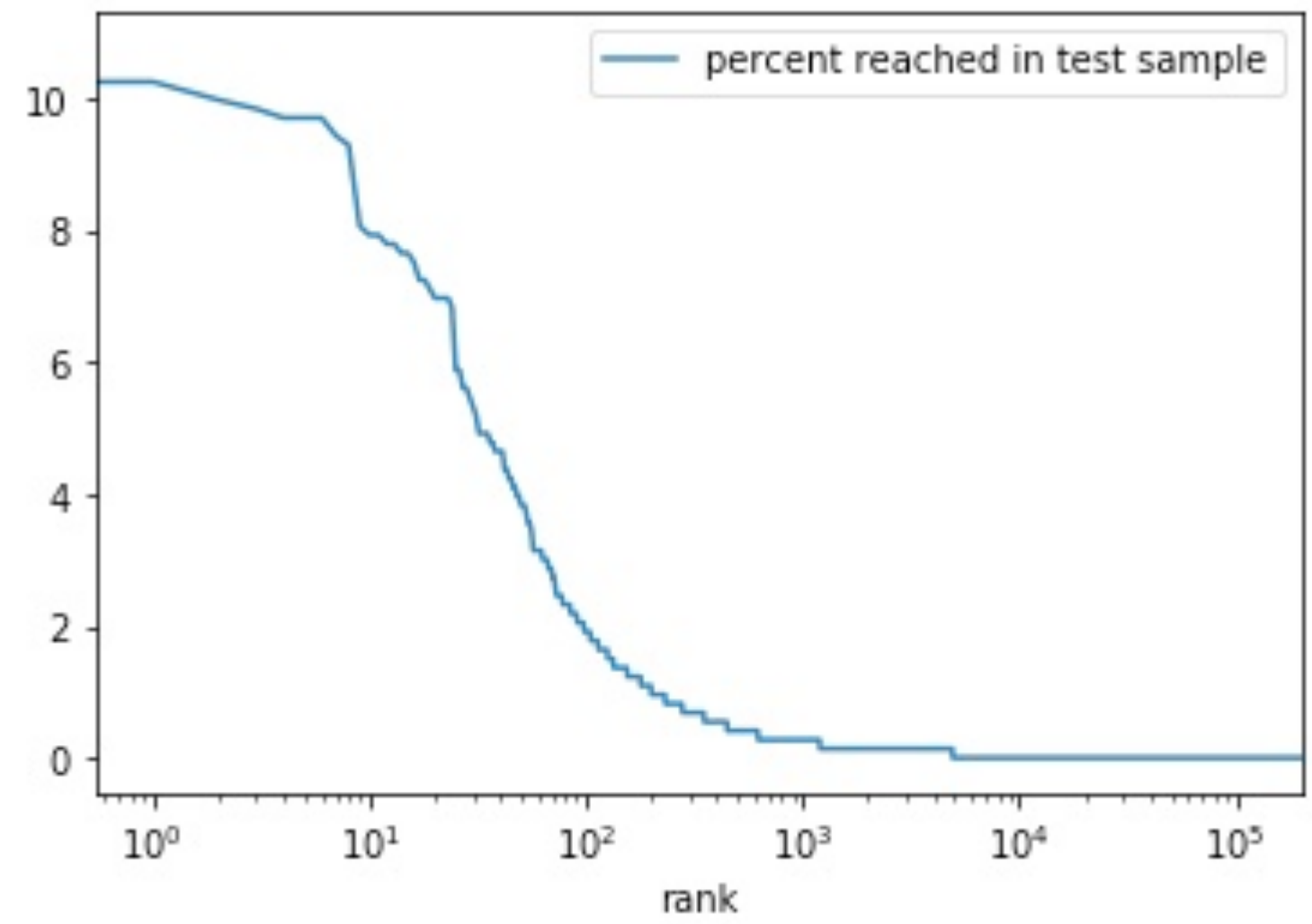

***Figure 10:*** *Rank-reach distribution of accounts in the influencer sample (filtered for in-degree >= 1, leaving 199,180 accounts)*

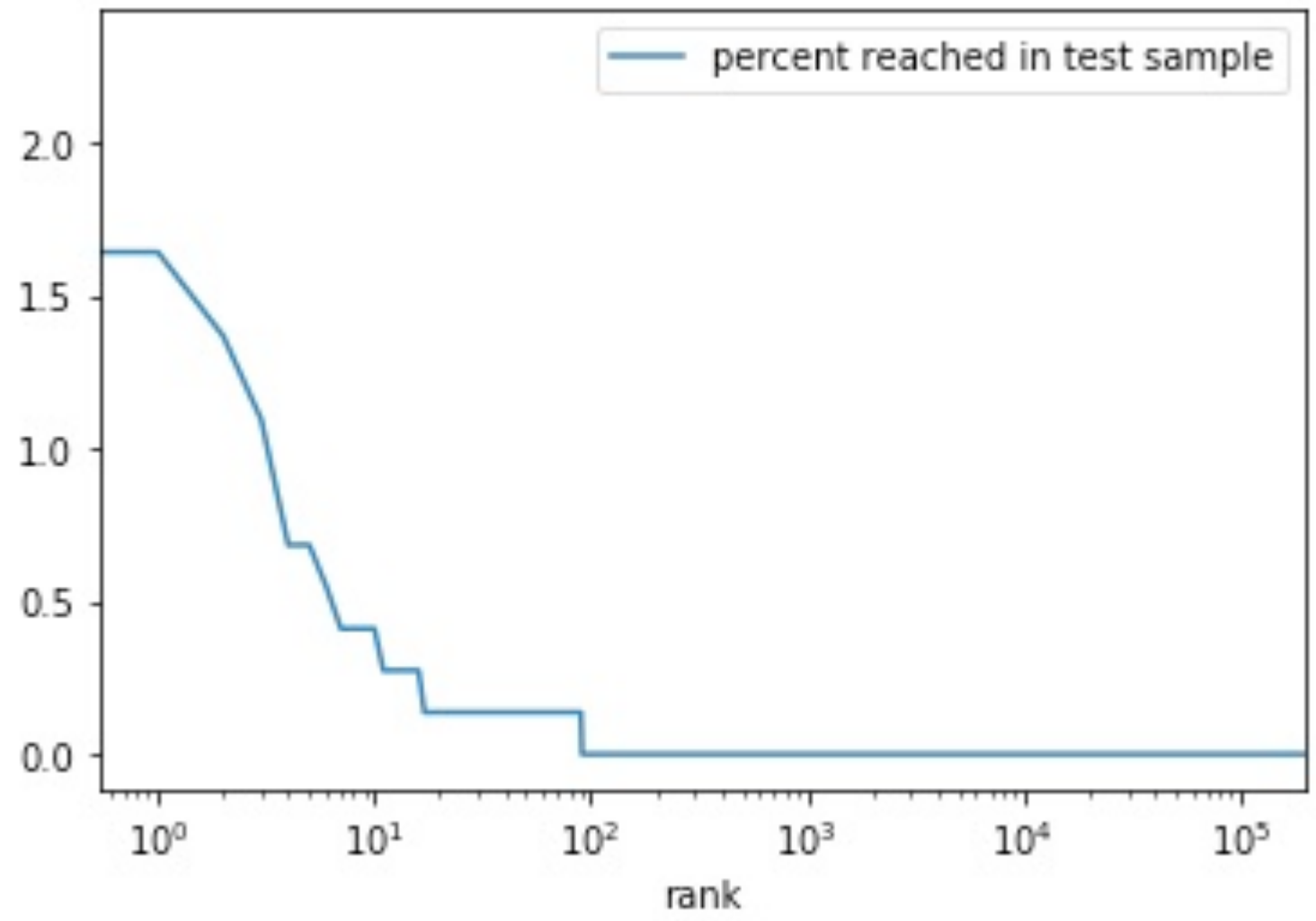

***Figure 11:*** *Rank-reach distribution of accounts in the baseline sample (199,180 accounts drawn randomly from German-using accounts in TrISMA collection)*

The categorical difference between our influencer sample and the baseline becomes even more clear when examining the rank-distributions of coverage: while Figure 8 shows a linear decline of coverage with rank for the influencer sample, Figure 9 illustrates that in the random baseline Twitter sample, coverage follows a seemingly exponential decline. The same holds true for the more intuitive concept of reach, i.e. the percentage of accounts in the test sample reached by accounts in the influencer and baseline sample, respectively. Here, Figure 10 and 11 show that while the top 10 accounts in the influencer sample each reach 8 to 10 percent of the test sample, not even the top account in the baseline sample reaches 2 percent in the test sample. In total, the influencer sample reaches 85 percent of the test sample accounts with more than 1 friend.

In summary, our influencer sample shows not only a class difference in activity and follower numbers compared with the average, but also contains on average 40 percent of the friends of a German-using Twitter account, and reaches 85 percent of accounts in the test sample with more than one friend. If we use 2.5 million weekly active accounts in Germany (Frees & Koch, 2018) as a conservative population estimate (instead of 15 million based on the TrISMA collection), our sample still represents less than 10 percent of this population. Taking this and everything above into account, we conclude that the influencer sample is a good approximation of the most influential core of the German-using Twittersphere.

### *Test Case: Topical Communities in the German Twittersphere*

To test the suitability of our adaptation of the rank degree method to investigate the overall structure of a language-based Twittersphere, we replicated an analysis of the full Australian Twitter follow network by Münch (2019, Chapter 6) with the 3-core of the full sample[10]. In the Australian case, this analysis combined community detection within the follow network and keyword extraction from the tweets of the respective communities to detect Twitter accounts with common topical interests and reveal the overall structure of the Australian Twittersphere. The filtering for the 3-core was done in order to avoid trivial star-shaped follow-back communities, which seem to be an artefact of the sampling method and affected the detection of useful communities. This filtering left us with a network of about 66,000 nodes and ca. 655,000 edges, i.e. less than 10 percent of the full network's accounts but over 40 percent of its edges. Consequently, it has to be noted again that this analysis focuses on the central core of influential accounts in the German Twittersphere and not on average German Twitter accounts.

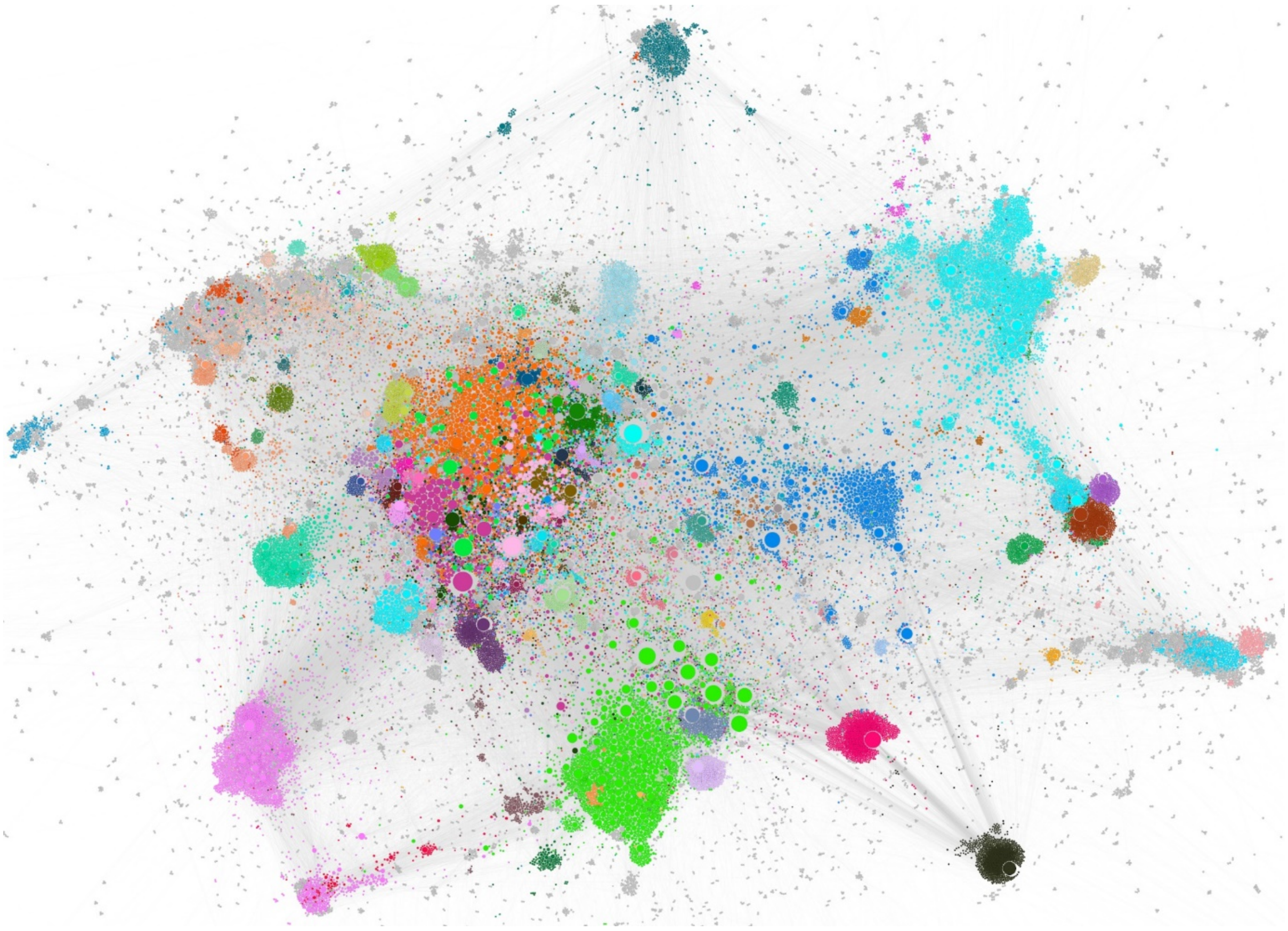

***Figure 12:*** *central communities in the 3-core of our sample network; coloured by largest communities detected with the infomap community detection algorithm; node size represents Page Rank (Brin & Page, 1998); layout done with Force Atlas 2 in Gephi (Bastian et al., 2009) (coloured version available online)*

[10] This includes the seeds. However, due to the 3-core constraint, seeds with no incoming edges will be excluded from this analysis here as well.

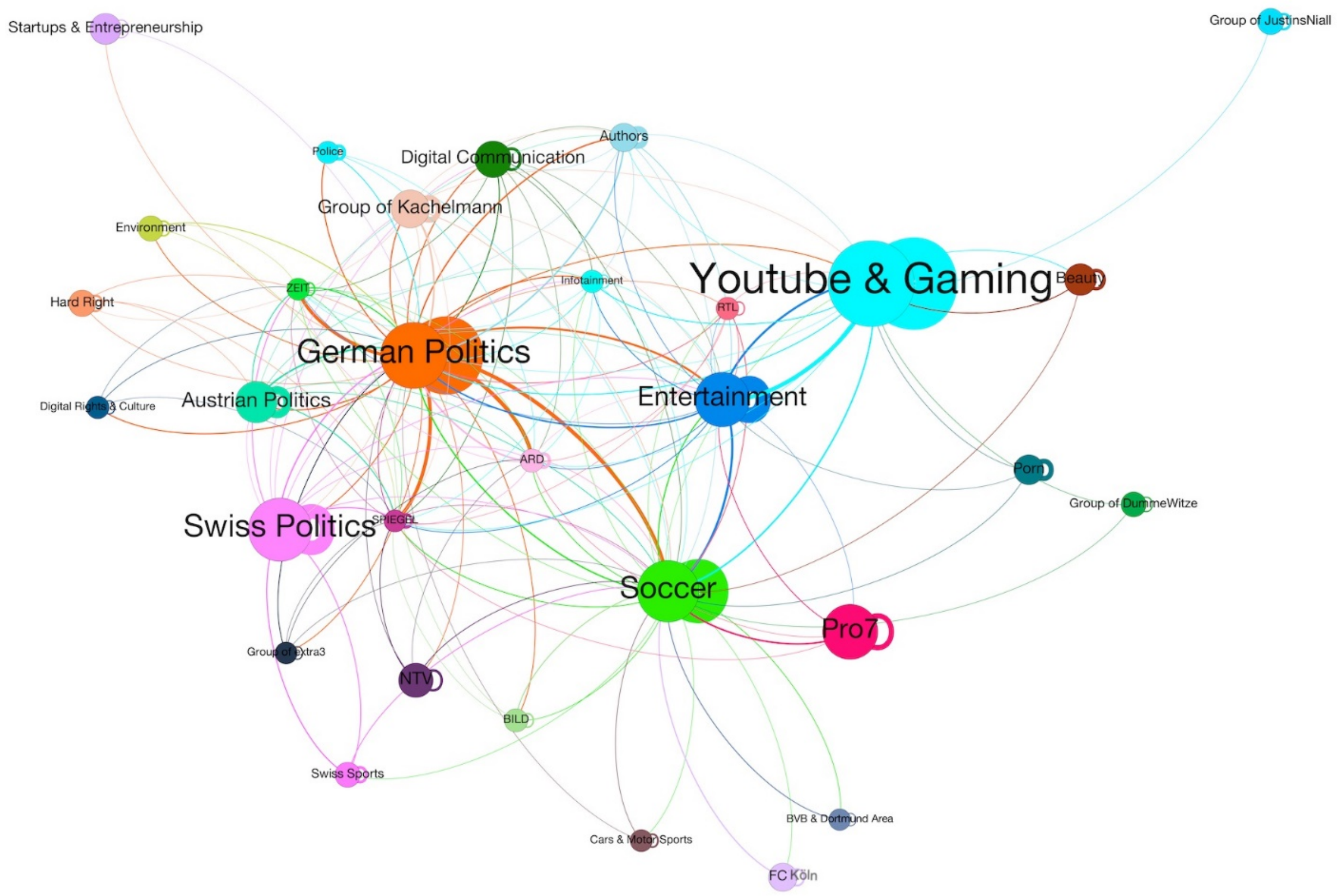

***Figure 13:*** *Community graph of communities in the 3-core of our sample with over 300 accounts, at least 80 active accounts during the examined time frame, and edges with a weight of at least 150; edge width represents weight; edge direction follows clockwise curvature; edges coloured by source node; node size represents the number of accounts in each community; node colours correspond with Figure 12; node labels based on interpretation of keywords and top accounts (see Supplemental Material); (coloured version available online)*

*Community Detection.* Instead of the Parallel Louvain Method (PLM) (Staudt et al., 2016), that was used by Münch (2019) and is based on modularity maximisation, ergo on a density-based understanding of community (Coscia et al., 2011), we used the non-hierarchical, non-overlapping version of the infomap algorithm (Rosvall et al., 2009; Rosvall & Bergstrom, 2008). This entropy-based algorithm is based on shortening the theoretical description length of the path of a random walker through the network by grouping nodes together. As a result, areas where a random walker would likely spend more time in a row are grouped together. In our case, if a tweet would be randomly shared along the network, it would on average stay within those communities for a longer time before leaving them. This intuitive interpretation and the fact that it allows, in contrast to the PLM, for a directed interpretation of the network, as well as other statistical advantages of the infomap method, lead to the decision to present its results instead of the results of the modularity maximisation based algorithm.[11]

[11] We also conducted the same analysis with the PLM algorithm. However, as expected, it resulted in worse results in the following keyword extraction. We used infomap with its standard setting of a Markov time of 1.

| **active accounts** | **Keywords** | **top accounts** | **tag** |
|---|---|---|---|
| 2015 | e3, stream, xd, e32019, nintendo, twitch, game, crossing, pc, zelda, animal, gameplay, cyberpunk2077, games, switch, xbox, trailer, cyberpunk, gaming, xboxe3, uff, nice, awesome, keanu, pk, live, nen, lol, mega | unge, dagibee, Gronkh, MelinaSophie, LeFloid, iBlali, Taddl, rewinside, HandIOfIBlood, PietSmiet | YouTubers & Gaming |
| 1855 | berlin, innen (female suffix), spd (German party), berliner, study, companies, discuss, cdu (party), demand, topics, important, federal government, german, digitisation, topic, has been, annefrank, climate protection, more, june, shows, germany, interview, politics, brandenburg | tazgezwitscher, Die_Gruenen, Tagesspiegel, c_lindner, gutjahr, dunjahayali, sigmargabriel, sixtus, HeikoMaas, spdde | German politics |
| 1414 | women's strike, switzerland, swiss people, glarner, svp (Swiss party), bern, women's strike2019, zurich, canton, grand, basel, national council, women | NZZ, 20min, viktorgiacobbo, Blickch, tagesanzeiger, srfnews, MikeMuellerLate, watson_news, migros, srf3 | Swiss politics / women's strike |
| 1044 | season, trainer, bundesliga, new arrival, player, dfb, fc, gerest, em, exchanges, estonia, exchange, transfer, goal, victory, team, wm, cup, liga | DFB_Team, FCBayern, ToniKroos, MarioGoetze, esmuellert_, Podolski10, Manuel_Neuer, Bundesliga_DE, ZDFsport, JB17Official | German football |
| 767 | övp (Austrian party), spö (Austrian party), fpö (Austrian party), vienna, austria, bierlein, oenr, austria's, ibiza, viennese, strache, kickl, turquoise, hofer, parliament, national council, zib2, abdullah, mandate, election campaign, centre, chancellor (female form), heinz, austrian, glyphosate, proposal, blue | florianklenk, sebastiankurz, IngridThurnher, kesslermichael, HannoSettele, vanderbellen, Gawhary, HBrandstaetter, HHumorlos, michelreimon | Austrian politics |
| 376 | dessau, roßlau, afd (German (far-)right-wing party), görlitz, islam, raped, migrants, asylum seeker, dangerous person, rejected person, sed, wippel, patriots, greta, niger, hosni, strongest, green, fridayforfuture, african, rosslau, left, gretathunberg, crime, old parties, merkel, radical left, greens, habeck, keep silent, tear apart, vote, girl, refugee, saxony, rape, maas, citizen, islamistic, sexual | DonJoschi, AfD, MSF_austria, Alice_Weidel, SteinbachErika, Joerg_Meuthen, Beatrix_vStorch, GrumpyMerkel, krone_at | hard right / xenophobia / migration / refugees |

***Table 2:*** *Selected communities' keywords (translated), top accounts by in-degree in the 3-core of our sample and our summarising tags.*

*Keyword Detection.* In order to determine topical keywords for the detected communities, we retrieved the last 200 tweets for every unprotected account in the 3-core of our sample. This dataset was filtered for tweets by accounts in the 93 communities with more than 100 accounts. Then, we filtered those for tweets posted in the last 7 days. Within this dataset, only 4.4 percent of the active accounts had tweeted more than 200 tweets. The collection took about 2 days. Therefore, to avoid having more tweets from accounts that were collected later

in the collection period, we cut off the last two days of these tweets. This left us with about 455,000 tweets by ca. 20,000 accounts over a period of 5 days (9.-14. June 2019).

The keyword detection process followed the same procedure as described by Münch (2019, p. 227), except for the use of German stop-words from the python-stop-words project[12] and the use of the unfiltered communities instead of their k-cores, due to the already filtered nature of the sample's 3-core. The keyword detection is based on the chi-squared statistic and is common in corpus linguistics (Rayson et al., 2004). The process returns a list of keywords ranked by how significantly being assigned to a group is correlated with the use of these keywords. We keep the top 50 keywords[13] and filter out keywords that have been used by less than 5 percent[14] of the respective community.

*Results.* Figure 12 shows the result of the community detection. On a first glance it becomes clear that the infomap algorithm in most cases still finds communities that align with the force-directed layout (done with Force Atlas 2 (Bastian et al., 2009)) – as would a modularity maximising algorithm. Already an inspection of the account names (the top 10 accounts by degree of communities with more than 100 active accounts within the 3-core are available in Appendix A) revealed that most of the largest communities have a topical focus.

This was confirmed by a close reading of the Twitter profiles of the top 10 accounts by degree in the 3-core and triangulated with an interpretation of the keyword analysis. As the account names, the keywords and the tags reflecting our interpretation of the communities with more than 100 active accounts in the analysed time period can be found in Appendix A. A selection in Table 2 demonstrates the topical clarity that enabled us to summarise the keywords and accounts to a topical tag. However, as the 'Hard Right' community demonstrates, it is important to stress that belonging to a community in this analysis does not necessarily mean endorsement of its majority's activities: While most of the top 10 accounts in this community can be identified as members of the German right-wing party AfD or accounts obviously supporting this party, we also find 'krone_at', the account of an Austrian tabloid, and 'MSF_austria', the account of Doctors Without Borders in Austria. For the latter, we could determine that this is likely due to the fact that prominent accounts in the 'Hard Right' cluster follow 'MSF_austria'. 'MSF_austria' does not follow them back.

Finally, the summary of this test study is depicted in the community graph in Figure 13 which contains the tags summarising our interpretation of the keywords and top accounts. If we could not find a clear interpretation, the community is tagged as 'Group of' the account with the highest degree in the 3-core. While many communities and their connections are filtered out for clarity and only the largest, most active communities and strongest connections between them remain, it gives a useful bird's-eye view on the structure of the analysed network, which, according to our results above, represents the influential core of the German Twittersphere. As such it exhibits intuitively sensible patterns at first sight: Swiss Politics is strongly connected with Swiss Sports and vice versa; Hard Right, and Digital Rights Culture appear as satellites of the dominant German Politics community; Porn is remote from most communities and follows more than it is followed; and Youtubers & Gamers are connected to the rest of the network mostly through Entertainment. In short, this result resembles the results for the Australian Twittersphere by Münch (2019, Chapter 6) and provides a good overview of the influential core of the German-speaking Twittersphere.

---

[12] https://pypi.org/project/stop-words/

[13] This is an arbitrary cut-off, mostly based on our estimate how many words a reader can capture to recognise common topics. Our results remained robust with significantly higher (>100) or lower (<10) numbers.

[14] Again, an arbitrary chosen percentage, based on our understanding of what we can still call a keyword. Also here, results stayed mainly the same in the later analysis with other thresholds, except for less tagged communities for higher and more tagged communities for lower thresholds.

# Conclusion

Summarising the methods and results described in detail in the section above,

- we have adapted the rank degree method in a way that makes it practically useful for a small team in order to gather Twitter follow networks of most influential accounts that use a certain interface language using the cost-free Twitter standard APIs;
- we provide evidence that a network sample collected with this method exhibits activity and follower numbers orders of magnitude higher than the average of this language-domain;
- we provide evidence that the influencer sample, i.e. a subsample of accounts with at least 1 incoming edge within our sample, represents under 10 percent (likely much less) of the whole population, but reaches 85 percent of accounts in a random test sample of German-using accounts with more than 1 friend;
- we show that accounts within the influencer sample exhibit substantially higher reach than accounts in a random baseline sample of the same size; and
- we provide evidence that for an average German account 40 percent of its friends are in our influencer sample; therefore, given the higher activity of the sample, likely more than 40 percent of an average German Twitter account's timeline is produced by our sample.

Altogether, this lets us conclude that the adapted sampling technique is able to approximate the set of most influential accounts, based on the follow-network, within a language-based Twittersphere. In comparison to the original rank degree method, our adaptation of the method is optimised to be parallelisable and efficient concerning API calls and can therefore be used by small research teams and social media professionals. Furthermore, we are confident that with some adaptations it is also suitable to mine not only Twitter follow networks but also comparable platforms that have a subscription network.

As our test study demonstrates, the data retrieved with this method enables a researcher to conduct research projects that hitherto relied on much larger datasets and data collections. Bruns et al. (2017), Bruns & Enli (2018), and Münch (2019), all relied on the full follow network that was collected based on the global Twitter account collection by TrISMA. While they filtered a complete dataset down to a manageable or useful set of likely influential accounts, our method restricts the collection to those accounts in the first place. It produces a comparatively small sample of connections in the follow network based on a random sample of accounts[15] leading to a backbone network of the most central, thus likely the most influential accounts. As a positive and important side effect, this also leads to less ethical issues, as it restricts the data collection mostly to accounts that are already popular on Twitter and therefore more likely to be aware of the public availability of their data.

Despite the smaller scale of the produced dataset, we were able to retrieve meaningful, comparable results to the three studies above, using and triangulating their methods. While the focus of this paper is on the sampling method and we do not dive deeper into the theoretical implications of the test study, it is clear that the test study presents a fertile ground for theory development comparable to Bruns & Highfield (2016), for example, or discourse analysis as done by Dehghan (2018). We want to especially highlight the observed similarities between our representation of the overall structure of the German Twittersphere and the Australian Twittersphere as drawn by Bruns et al. (2017) and Münch (2019, pp. 237–238), which hints at an untapped potential for international comparative communication and (social) media studies.

[15] While we used all 15 million accounts that used German as interface language in the TrISMA dataset, a smaller sample also would have sufficed as a seed pool.

# Outlook

This study provides evidence that our adaptation of the rank degree method enables drawing a representative language-based sample of influential Twitter accounts. However, while we are confident that the method collects the most influential accounts despite our adaptations, for example, the exact ranking of these accounts by different centrality measures, as well as community structures might be differently preserved than in the original method. Therefore, this method still has to be tested with known networks, as the Australian or Norwegian Twittersphere, to ensure that the centrality-preserving qualities of the original rank degree method do not suffer from our adaptations. Especially the application to a directed network, as well as the non-dynamic handling of degree and ranking in the sampling process might lead to significant differences in the sample quality.

Nevertheless, now that the practical feasibility of the method is proven, the quality of the sample can be assessed more thoroughly, especially regarding the question how coverage and reach change with the sample size – a most important question for researchers and social media professionals.

Moreover, for the current version of the method we still required a high number of initial random seeds for the algorithm to work, and this seed collection is generally not given. Thus, a further development of our prototype includes the generation and growth of the seed pool. Ideas include sampling via a keyword search for common words in the language or regarding the topic of interest and a ‘snowballing’ approach, where the latest 5000 connections of the collected nodes are stored as seeds[16]. Such an implementation needs to be tested for representativity and comparability with the current approach, as the seed pool itself is not random anymore (Münch & Rossi, 2020).

As Twitter made the interface language of an account a private property at the end of this project, the method has been adapted to work with the tweet language instead. Whether the language detection provided by Twitter suffices for this remains to be tested.

Finally, further avenues of enquiry regarding this form of sampling include the collection and comparison of language-based Twitterspheres other than German, and its development to social media mining approaches that are based on topical instead of language-based criteria (Münch & Thies, 2020).

## Acknowledgements

We want to acknowledge the postdoctoral research network Algorithmed Public Spheres at the Leibniz Institute for Media Research | Hans-Bredow-Institut (HBI), the TrISMA project, and our contributing colleagues at the HBI, namely Wiebke Loosen, Christiane Matzen, Jan-Hinrik Schmidt, and Johanna Sebauer, for their trust, interest, and support for the collaborative data collection effort needed to realise this project.

## Declaration of Conflicting Interests

The authors declare that there is no conflict of interest.

## Funding

The author(s) received no financial support for the research, authorship, and/or publication of this article.

[16] We retrieve this information already for selecting the highest ranking friend of a node.

# Author biographies

Felix Victor Münch works as a Postdoc at the Social Media Observatory of the Leibniz Institute for Media Research | Hans-Bredow-Institut in Hamburg. With a PhD from the Digital Media Research Centre at QUT (Brisbane, Australia), an M.A. in Journalism (LMU and German Journalist School, Munich, Germany), a B.Sc. in Physics (LMU, Munich, Germany), and work experience in online media brand communication as an online media concepter, user experience designer, and strategist, he is most likely a computational social scientist by now. At the moment he focusses on network science, social media, and theories regarding the public sphere.

Ben Thies (BA, Zeppelin University Friedrichshafen) is a graduate student of statistics at Humboldt University Berlin, Free University Berlin, and Berlin Institute of Technology. During this study, he was a research assistant at the Leibniz Institute for Media Research | Hans-Bredow-Institut in Hamburg. Now he works at the Mercator Research Institute on Global Commons and Climate Change. His main interests lie in human behaviour, online social networks, and Bayesian networks.

Cornelius Puschmann is professor of media and communication at ZeMKI, University of Bremen and an affiliate researcher at the Leibniz Institute for Media Research, as well as the author of a popular German-language introduction to content analysis with R. His interests include digital media usage, online aggression, the role of algorithms for the selection of media content, and automated content analysis.

Prof. Axel Bruns is a Professor in the Digital Media Research Centre at Queensland University of Technology in Brisbane, Australia, and a Chief Investigator in the ARC Centre of Excellence for Automated Decision-Making and Society. His books include Are Filter Bubbles Real? (2019) and Gatewatching and News Curation: Journalism, Social Media, and the Public Sphere (2018). Bruns served as President of the Association of Internet Researchers in 2017–19. His research blog is at http://snurb.info/, and he tweets at @snurb_dot_info.

# Appendix

## Keywords and Top Accounts in Detected Communities of Sample 3-Core with Over 100 Active Accounts

| *active accounts* | *Keywords* | *top accounts* | *tag* | *short tag* |
|---|---|---|---|---|
| 2015 | e3, stream, xd, e32019, nintendo, twitch, game, crossing, pc, zelda, animal, gameplay, cyberpunk2077, games, switch, xbox, trailer, cyberpunk, gaming, xboxe3, uff, nice, geil, keanu, pk, spiele, live, nen, lol, mega | unge, dagibee, Gronkh, MelinaSophie, LeFloid, iBlali, Taddl, rewinside, HandIOfIBlood, PietSmiet | YouTubers & Gaming | Youtube & Games |
| 1855 | berlin, innen, spd, berliner, studie, unternehmen, diskutieren, cdu, fordern, themen, wichtiger, bundesregierung, deutschen, digitalisierung, thema, wurde, wurden, annefrank, klimaschutz, mehr, juni, zeigt, deutschland, interview, politik, brandenburg | tazgezwitscher, Die_Gruenen, Tagesspiegel, c_lindner, gutjahr, dunjahayali, sigmargabriel, sixtus, HeikoMaas, spdde | German politics | German Politics |
| 1414 | frauenstreik, schweiz, schweizer, glarner, svp, bern, frauenstreik2019, zürich, kanton, grosse, basel, nationalrat, frauen | NZZ, 20min, viktorgiacobbo, Blickch, tagesanzeiger, srfnews, MikeMuellerLate, watson_news, migros, srf3 | Swiss politics / women's strike | Swiss Politics |
| 1044 | saison, trainer, bundesliga, neuzugang, spieler, dfb, fc, gerest, em, wechselt, estland, wechsel, transfer, tor, sieg, mannschaft, wm, cup, liga | DFB_Team, FCBayern, ToniKroos, MarioGoetze, esmuellert_, Podolski10, Manuel_Neuer, Bundesliga_DE, ZDFsport, JB17Official | German football | Soccer |
| 767 | övp, spö, fpö, wien, österreich, bierlein, oenr, österreichs, ibiza, wiener, strache, kickl, türkis, hofer, parlament, nationalrat, zib2, abdullah, mandat, wahlkampf, zentrum, bundeskanzlerin, heinz, österreicher, glyphosat, antrag, blau | florianklenk, sebastiankurz, IngridThurnher, kesslermichael, HannoSettele, vanderbellen, Gawhary, HBrandstaetter, HHumorlos, michelreimon | Austrian politics | Austrian Politics |
| 751 | menschen, nazi, merz, deutschland, afd, wer, akk, cdu, spd, jahre, leben, wäre, grünen, obwohl, jemand, klar, nazis, ja, liebe, wähler, bitte, viele, immer, mann, wissen, wünsche, herr, seit | ntvde, Muhterem_Aras, munich_startup, Literaturtest, _schwarzeKatze, bmzimmermann, ninjawarriorrtl, cesurmilusoy, MetalabVie, S_chill_ing | NTV (news channel) | NTV |
| 720 | album, Keanu | officiallyjoko, damitdasklaas, siggismallz, thisiscro, latenightberlin, gamescom, EtienneToGo, ralphruthe, TheRocketBeans | entertainment | Entertainme nt |
| 563 | | saschalobo, StefanOsswald, CarstenRossiKR, ragnarh, cbgreenwood, breitenbach, BBDO, Ugugu, ring2, julianheck | Digital Communication | DigiCom |
| 510 | | Kachelmann, schmitt_it, RomeoMicev, Perspektive360, paulespcforumde, naturkosmetix, TUIDeutschland, glockendoktor, nachtnebel, HaraldKlein | ? | ? |
| 453 | lt, bett, morgens, hast, schön, hasse, müde, essen, ja, möchte, manchmal, mama, bitte, frau, nachts, immer, bier, nein, kaffee, gern, gar, jemand, weiß, mal, geh, glaube, ach, scheiße, tweets, hund, kinder, hätte, abends | KuttnerSarah, Regendelfin, vergraemer, DrWaumiau, HappySchnitzel, katjaberlin, ArminRohde, RenateBergmann, hashcrap, peterbreuer | "Feuilleton" / authors / copy Writers | Authors |
| 429 | 10jahrejk, joko, klaas, mtv, canadiangp, hitzekindofmagic, fernsehen | ProSieben, MattBannert, newshausen, haveonenme, Mone_Horan, about_riki_, delay1, sarahofer8, sweettweetangie, JoanBleicher | Pro7 (Private TV broadcaster) | Pro7 |
| 376 | dessau, roßlau, afd, görlitz, islam, vergewaltigt, migranten, asylbewerber, gefährder, abgelehnter, sed, wippel, patrioten, greta, niger, hosni, stärkste, grüne, fridayforfuture, afrikaner, rosslau, linke, gretathunberg, asylbewerbe, verbrechen, altparteien, merkel, linksradikale, grünen, habeck, schweigt, zerreißen, wählt, mädchen, flüchtling, sachsen, vergewaltigung, maas, bürger, islamistischer, sexuell | DonJoschi, AfD, MSF_austria, Alice_Weidel, SteinbachErika, Joerg_Meuthen, Beatrix_vStorch, GrumpyMerkel, krone_at | hard rights / migration / refugees | Hard Right |
| 369 | mydirtyhobby, femdom, boobs, latex, milf, mistress, cock, anal, tits, blowjob, busty, fetish, livecam, pornstar, mdh, sexy, findom, highheels, clip, manyvids, cam, cumshot, cum, webcam, pussy, goddess, dirty, bdsm, porn, hot, horny, booty, slave | Erotik_Center, AnnyAuroraPorn, texas_patti, LenaNitro1, sandy226, Julietta_com, ModelRoxxyX, swo2212, LucyCatOfficial, SuziAnneVX | Porn | Porn |
| 275 | ariana, literally, 5sos, love, stan, heart, she, me, omg, wanna, deserve, him, lmao, them, tattooed, ily, please, like, my, someone, selena, know, cutest, much, being, gonna, really, don, cant, life, when, music, but, thank, ever, miss, got, look, fucking, happy, do, they, did, said, can, re, ll | JustinsNiall, ifuckzayn, o2de, PlanetLouis, rickthejizzler, infernelle, betteoroff, marvelpurpose, wanderlxsthes, Markus_18 | ? | ? |
| 261 | verkehrsunfall, hinweise, polizei, fahndung, zeugen, einsatz, pkw, unbekannte, kollegen, verletzt | PolizeiMuenchen, PolizeiBerlin_E, Polizei_Ffm, PolizeiHamburg, BMI_Bund, PolizeiSachsen, bka, Berliner_Fw, polizei_nrw_k, PP_Stuttgart | Police | Police |
| 256 | Effzeh | fckoeln, michelpauwels, _AlienAlliance, saschwelt, skopyM_, colognecitynet, KoelnFormat, MadlenKaniuth, amendedestages, marketingshopde | soccer club FC Köln/Cologne area | FC Köln |
| 248 | bpshutdown, ölplattform, greenpeace, bp, klimakrise, aktivist, rwe, climate, klimaschutz, landwirtschaft | WWF_Deutschland, greenpeace_de, stARTconference, SMAsolar, energynet, tp_da, globalmagazin, 100ProzentEE, GreenpeaceAT, global2000 | Environmentalism | Environment |
| 246 | moin, lieben, follower, schön, wünsche, bitte, schönen, cdu, containern, menschen, gut, liebe | Wurfschuh, DeutscheOnline, mauerunkraut, DieLinkeBremen, twetter, EventsbyWW, KlatschKopp, emet_news_press, OBuerge, Mama_notes | ? | ? |
| 240 | wdr, übergangsweise, wieben, wilhelm, männlicher, ard | tagesschau, WDR, DasErste, tagesthemen, tagesschau_eil, ARD_Presse, ARDde, 3sat, ndr, Tatort | Public Service Broadcasters ARD | ARD |
| 238 | | janboehm, PixelHELPER, UensalArik, DHallervorden, schmarsten, tweetbarth, BruneKerstin, hwieduwilt, BuzzFeedNewsDE, foodoraboi | Comedy / Political Satire / Infotainment | Infotainment |
| 236 | Privacy | netzpolitik, digiges, lorz, friiyo, leonidobusch, bendrath, andre_meister, netzpolitik_org, tmrazek, peterschink | digital rights and digital culture | Digital Rights & Culture |
| 215 | chernyshev, mfa, detained, ausgewiesen, golunow, producer, moskau | SPIEGELONLINE, SPIEGEL_EIL, DerSPIEGEL, SPIEGEL_Top, SPIEGEL_Politik, SPIEGEL_Sport, SPIEGEL_Netz, SPIEGEL_Wissen, SPIEGEL_Kultur, SPIEGEL_alles | SPIEGEL (German News Magazine) | SPIEGEL |
| 206 | | extra3, uiuiui7, tibor, Zellmi, jugendmedien, Piratenbaer, | ? | ? |
| | | deuxcvsix, msulzbacher, annikrubens, L_Petersdotter | | |
| 202 | | zeitonline, RegSprecher, DIEZEIT, zeitonline_pol, zeitonline_kul, zeitonline_dig, ZEITmagazin, berndulrich, zeitonline_wis, zeitonline_wir | DIE ZEIT (weekly broadsheet newspaper and online news) | ZEIT |
| 186 | bvb, isak, zorc, gündogan, dortmund, borussia, reus | BVB, stadtdortmund, MauriciusQ, WR_Dortmund, fluestertweets, DahlHolger, DortmundBlog, Ina_Steinbach, RN_Florian, PottblogLive | Borussia Dortmund (soccer club) & Dortmund area | BVB & Dortmund Area |
| 184 | Goalie | blickamabend, srfsport, XS_11official, swissteam, SteffiBuchli, SwissIceHockey, SFV_ASF, GokhanInler, JOIZTV, FCBasel1893 | Swiss Sports | Swiss Sports |
| 172 | Rtl | RTLde, DWDL, rtl2, kabeleins, serienjunkies, promibb, voxdhdl, DMAX_TV, burda_news, Sila_Sahin | RTL (private TV broadcaster) | RTL |
| 171 | jungkook, bts, jimin, yoongi, 6yearswithourhomebts, taehyung, namjoon, 방탄생일ㅊㅋ, armys, 축하해, 방탄소년단, 2019btsfesta, 우리, 방탄6주년보라해, 저도, 주셔서, 작은, 선물, 준비했어요, 방탄생일, 감사해요, 벌써, 여행, hbd_bts, 6년, 부산아이가, seokjin, 알라뷰, 하트, 여섯애기들, 생일, 축하행, 아미도, 석지니와, festa, jin, hoseok, jk, 살랑살랑, 꼬리, 6주년, euphoria, army, tae, 방탄다락, bangtan, 613, 가께, 기달려, 형이 | btswordwide, meteorbts, ThePhilCoenen, TattedxUp, moonrosecom, BTSGIobal, pradakookie, BieberHades, MartinaOppel, TOTALLYBIA | ? | ? |
| 170 | lemans24, wec, qualifying, dtm, bmw, lemans, race, motorsport, car, mans, porsche, drivers | HulkHulkenberg, PorscheRaces, BMWGroup, realTimoGlock, PorscheNewsroom, autobild, nuerburgring, NickHeidfeld, MB_Museum, ADAC | cars & motor sports | Cars & Motor Sports |
| 152 | | SZ, SZ_TopNews, SZ_Muenchen, SZ_Sport, SZ_Kultur, SZ_Digital, SZ_Bayern, kopfzeiler, SZ_Karriere, SZ_Gesellschaft | Süddeutsche Zeitung (broadsheet newspaper and online news) | SZ |
| 150 | bir, çok, olsun, için, gibi, kadar, bu, güzel, yok, daha, değil, hiç, ama, bana, sonra, ile, olan, diye, benim, iyi, kendi, biz, türkiye, yalan, sana, başka, gün, beni, oldu, stanbul, siz, destek, diyor, artık, allah, türk, tüm, mı, var, şey, tek, önce, bile, chp, nasıl, sizin, ın, yine, ye, yeni | sandukankanack, ElLobo0815, akbasmarkt, AbdullahSOnal, romanticfilozof, togisitompul, truppenn, Der_Turken, PeterPollak1 | Turkish | Turkish |
| 145 | multilateralismmatters, ghanaian, chilean, ministers, ghana, disarmament, chile, multilateralism, alemania, amb, foreign, ahora, cooperation | AuswaertigesAmt, GermanyDiplo, PaulNemitz, LN_Online, wahl_de, amtzweinull, AvHStiftung, dw_politics, ResoluteSupport, dgapev | international relations, foreign affairs, policy | Foreign Affairs |
| 142 | bildplus, inhalt | BILD, BILD_News, BILD_Promis, Fussball_Bild, BILDamSONNTAG, BILDhilft, BILD_Berlin, BILD_Politik, BILD_Digital, BILD_TopNews | Bild (German tabloid) | BILD |
| 141 | cod, fortnite, bro, digga | MarcelScorpion, ViscaBarcaCoD, GunEliteCoD, KamoLRF, ELoTRiX, DC_Haptic, KiviCoD, AmarCoDTV, JamarzMW, DonRaven | Gaming | |
| 139 | bvg, jelbi, verspätungen, ausfällen | BVG_Kampagne, DVBAG, VBB_BerlinBB, BVG_Bus, | Berlin area | Berlin |
| | | BVG_Tram, VVS, RegineGuenther, Schienenallianz, VIZ_Berlin, BerlinPartner | | |
| 136 | Saarbrücken | peteraltmaier, JuergenHerzog, Truppenursel, henrikMSL, drmfuchs, OlliLuksic, Riotbuddha, Arnd_Diringer, HoneckerMargot, Enigma424 | ? | ? |
| 130 | | niggi, maithi_nk, tobiasgillen, Christiane, MikeSchnoor, ankegroener, tomhillenbrand, Wieprecht, frankzimmer, wortfeld | (media) journalists / authors | (Media) Journalists / Authors |
| 130 | | SBahnBerlin, Microstrom, SebastianHass, vomitalloverme, kanzleireiss, FilmTeam, sparbon, arschfax, Partnerschaft10, Kinderwunschnet | ? | ? |
| 126 | hvv, harburg, wilhelmsburg | mopo, hamburg_de, abendblatt, Coachforyou, hochbahn, SBahnHamburg, mein_hamburg, BILD_Hamburg, HVVStoerungen, Radio_Hamburg | Hamburg area | Hamburg |
| 123 | | ZDF, ZDFheute, ClausKleber, heutejournal, ReporterZDF, berlindirekt, _Helene_Fischer, dominikrzepka, ethevessen, ZDFinfo | Public Service Broadcaster ZDF | ZDF |
| 123 | marlies, moin, lieben, genieße, wünsche, schönen, wunderbares, wunderschönes | brettermacher, 1Wampel, LolainPause, emilyrosefrankT, OBIKO_de, hanne_fs, handzahm, keinlemming, schreibtatze | ? | ? |
| 119 | htc, samsung, xiaomi, pixel, oneplus, pixel4, u19e, 9t, galaxy, 3a, s10, xcloud, huawei, telefónica, xl, ios, asus | caschy, SamsungDE, ANDROIDPIT, Intel_DE, mobiFlip, pcwelt, LGDeutschland, connect_de, HTC_de, Technikfaultier | consumer technology / mobile phones | mobile |
| 117 | heartbroken, burns | welt, WELTnews, Schuldensuehner, omichalsky, WELT_Kultur, WELT_Economy, axelspringer, Nisalahe, RomanusOtte, Christian_Meier | WELT / Axel Springer | WELT |
| 116 | dokomi, manga, cosplayer, cosplay, loot, detektiv, anime, prints | carlsen_verlag, Animexx, KazeDeutschland, Connichi, Reyhans_Art, PepperAnime, TOKYOPOP_GmbH, ConanNews, DoKomi, AnimagiCtweet | Anime / Manga | Anime / Manga |
| 116 | | gruenderszene, Akkumonster, ralphdietze, mekofactory, Mobilstand, _mv, gutefrage_net, Leonie_Markus, otterstein, zeitreisender | startups and entrepreneurship | Startups & Entrepreneurship |
| 114 | Altparteien | tanjaplayner, balleryna, ModernArtLove, art_collector_a, Artlover_art, ModernArtFair, Shopalooo, startup_2020, UnitedCrowd_com | ? | ? |
| 114 | | DummeWitze, DoertheBi, LlONTTV, paauul18, sxrahbca, dailywithsina, lasernica, hey_justlisa, CoolesBienchen, greedyjessi | ? | ? |
| 113 | fplus, jungunternehmer, verhielten, juraprofessor, auslandsstudium, umfragehoch | faznet, FAZ_Feuilleton, FAZ_Politik, FAZ_Wirtschaft, FAZ_Eil, FAZ_NET, FAZ_Sport, FAZ_RheinMain, FAZ_Wissen, _donalphonso | FAZ (broadsheet newspaper) | FAZ |
| 113 | Handelsblatt | handelsblatt, hb_finanzen, DrKissler, MDowideit, StN_Blaulicht, lokfuehrer_tim, sphericon, ThomasKuhn, hb_technologie, Niebisch | Handelsblatt (economy & finance focused newspaper) | Handelsblatt |
| 107 | frankfurt, jpmcc, ffm | Stadt_FFM, gitarra, markussekulla, Twittwoch, paulinepauline, clab, Herusche, hackr, wmfra, DerLachmann | Frankfurt area | Frankfurt |
| 103 | Duisburg | WAZ_Redaktion, Duesseldorf, duisburg_de, Regio_NRW, artduesseldorf, wznewsline, DeinNRW, bastei_luebbe, 2muchin4mation, theaterdortmund | North Rhine-Westphalia | NRW |
| 98 | dienstjahren, holschneider, digitalnewsreport, agenturjournal, cvd, dpareporter, gründungsdatum, volbracht, bindeglied, parteipolitischen, objektiven, arbeitete, dpa | dpa, DirkKirchberg, ralfdrescher, TobiasRees, KNA_Redaktion, radioffn, news_de, burkhardewert, FunkeBerlin, Weilmeier | | |
| 95 | airlines, ryanair, boeing, eurowings, airbus, zrh, airways, lufthansa, avgeek, flights, flight, germania | airberlin, eurowings, TUIflyGermany, Airport_FRA, berlinairport, zrh_airport, HamburgAirport, ITB_Berlin, flomeint | | |
| 94 | easycreditbbl, playoffsbaby, baskets | deutschetelekom, ProSiebenMAXX, easyCreditBBL, TelekomBaskets, albaberlin, VideoMasterplan, Teichreport, BroseBamberg, skyliners1999, REWE_Supermarkt | | |
| 91 | | 1LIVE, RiberyFranck, SchornShow, HerrBursche, ThorstenIsing, onkelfisch, michaelimhof, johannataenzer, jczeller, flobraun | | |
| 90 | angriffslustig, habeck | RolandTichy, inpressulum, Hobbbes, vera_lengsfeld, rasenspiesser, SueWestCom, Unser_Luebeck, franz_rother, annatab, mhb_sport | | |
| 86 | wiwo | wiwo, Schnutinger, weinreporter, PRKanzlei, HeikoKunert, pr_ip, rrho, UNICUMde, astreim, GerdKotoll | | |
| 86 | remix, np, vinyl | berghain, PanPotOFFICIAL, meltfestival, BenKlock, TOCADISCO, cosmicgate, OstgutTon, kontorrecords, Kaiserdisco, Groove_Mag | | |
| 86 | | BibisBeauty, julienco_, xjolivex, glossybox_de, davyx21, lme704, DieBeautyGurus, JoKranz, nessaheil, LenooLe | Beauty | Beauty |
| 82 | | MartinSchulz, bayern2, UlrikeRodust, SPDThueringen, berndlange, ostkurve, SPDSachsen, OECDBerlin, SPDSH, JuttaSteinruck | | |
| 76 | | sternde, WN_Redaktion, ralfklassen, frankzdeluxe, LousyPennies, ByteFM, hmtillack, mesale_tolu, bzfe_de, vonstreit | | |
| 75 | | OomenBerlin, Bibliothomas, ZIMMER751, TunyBeats, soophiia_98, ellensohndavid, pbruegger, aminajxx, ekbo_de, WegenerHenrik | | |
| 71 | | rponline, rpo_duesseldorf, rpo_topnews, rpo_politik, rpo_sport, | | |
| | | Pillendreher, autobildtuning, kukksi, ReneRuebner, alextroll | | |
| 69 | energieeffizienz, jungtiere, luchse, fossile, wärmewende, steigend, energieträger, erneuerbareenergien | bmu, IMMOVATION_AG, StefWenzel, VollmarWWF, rischwasu, Dritter_Planet, HubertWeiger, EANRW, GEOMAR_de, Roland_Gramling | | |
| 66 | | stbaasch, mmbuelow, piixel_de, travellersknow, KlemensSkibicki, teuthorn, andiheer, birdy1976, princessCH, sue_jil | | |
| 66 | | PETADeutschland, heilerschule, oPetition, zukunftswerk, bourbonxirwin, sbstnbckhs, FORUM_CSR, mona__138, Joerg_Howe, Maureen_aha | | |
| 66 | xmen, kinostart, eiskönigin, paragraph175, meninblackinternational | MSchweighoefer, Elyas_MBarek, JanJosefLiefers, sixxTV, berlinale, k_herfurth, bambi, TilSchweiger, VeronicaFerres, BoraDagtekin | | |
| 65 | | alextv, alexibexi, iNetBlog, mikefinn_tweets, dts_nachrichten, bbswiss, farbwolke, Heinzus, Andivista, Flausenkopf | | |
| 60 | | wuv, Parallelwelten, online_praesent, egovcomde, SoMePraxis, salebyweb, reizwert, Tweetorator, partymusik24 | | |
| 59 | del2, kruminsch, arturs, überstandener, saarinen, sekesi, czarnik, pompei, willkommenzurück, horava, litesov, abstreiter, trainerausbildungen, starbulls, huba, ecn, pageau, esv, mathieu, shl, kaufbeuren, sokolov, krebsbehandlung, teilnehmerzahlen, johansson, deb, fenix, ravensburg, landshut, doppelpack, robbie | deb_teams, Koelner_Haie_72, Eisbaeren_B, adlermannheim, Ice_Tigers, EishockeyN, RedBullMuenchen, grizzlys_wob, SERCWildWings, roosters_hockey | | |
| 56 | | Afelia, 688i, AndiPopp, mueslikind, elawprof, TeilerDoehrden, SebastianLoudon, balu, ArminNassehi, Nien0r | | |
| 56 | vorverkaufsstart, jubiläumsfestivals, presale | eventimDE, rockamring, dietotenhosen, Wacken, hurricanefstvl, donotsofficial, metalhammer_de, feinesahne, BRLRS, rockimpark_com | | |
| 53 | schadsoftware | CHIP_online, Tweedmagazin, AllAboutPlus, ComputerWissen, PanasonicDE, IFA_Berlin, Fujitsu_DE, trendsblog, macwelt, magix_DE | | |
| 47 | | depechemode_de, pitsbiz, Staubkind_Music, gegisa, Mindbreed, kontaktperle_de, KornFlash, Koelnerbuerger, tischnachbar, sabrinadrums | | |
| 46 | | besserwerber, HNA_online, Liederpoet, charleemelody, superclix, 1apo, 997iit, Maler_Franchise, Suntmarkt, MicroTarife | | |
| 44 | cunha, sehnde, gulácsi, leistenverletzung, péter, dierotenbullen, astonvilla, aces, burgdorf, rbl, kaderplanung, naod | HAZ, DieRotenBullen, PhantomriderAMG, uestra, regionhannover_, kickeresport, sn_online, dpalni, DBRegio_NDS, kicker_RBL | | |
| 36 | popkünstler, gustaf, máxima, cumbria | Lenas_view, BUNTE, gala, OKMagazin, sarahconnorfc, HenningWehland, DKMS_de, carlottatruman, ENERGY_HAMBURG, Konni | | |
| 36 | | MarcelHirscher, anna_veith, dominiquegisin, dariocologna, TinaWeirather, carlo_janka, skiweltcup_tv, MorgensternT, BeatFeuz, michellegisin | | |
| 34 | iboprufen | TheRealLiont, tinaneumann_, _sxrx_tj, easiermutual, DanniKp, ce1inasc, Pia812002, lena_fichtner, lottixrowland, xIsabelx_x | | |
| 33 | economical | InStyleGermany, EbruErguener, GLAMOUR_DE, funnypilgrim, ELLE_de, essence_DE, TineValeska, freundin_de, rebeccafloeter | | |
| 27 | craaaaaazy, pilatus, viu, shegotme, hänni, hometown, mau | haenni__luca, Djantoine, JesseRitch, DJDave202, life140forever, shfanclub, tilllate_com, lealumusic, djginog, Tanja20114 | | |
| 25 | THE, leaned, WITH, SHIPS, THEMES, DETAIL, HEARD, DAUGHTER, **the**, MULTIVERSE, LITERATE, narrowed, TWILIGHT, WAS, HERE, RP, CHIEF, MATURE, rubs, MATCHES, **to**, DECSRIPTIVE, sighed, slits, grin, sls, gaze, **and**, kissed | AmatoryLissome, Arcanewildfire, DestroyedShadow, DamonGSalvatcre, TheOnlySexyJay, SwankyRadiance, BadWitchy, RPNetworkAgency, PenOfVoluptuous, diffidentminx | | |
| 23 | | sat1, buy_art_buy_art, cucky, malerdeck, GeldmitStrom, technikblog2019, jovisvisions, badbugs_art, lukes_woche, hochzeitskarten | | |
| 20 | | OneDirectionGER, FansOfCyrus, divazarry, travels_wonder, deadlytragedy, sunflowershawn, hollydreads, Katharina_Zim, Austin_Support_, jbiebur_ | | |
| 15 | philippians, artisans, surpr, unsettles, perseveres, incredib, thessalonians, callousness, unashamedly, unangezogen, sundayy, sowers, legalists, dailyprophetic, lukew, blameless, churning, nkjv, suaver, dashed, prophetic, abundant, amids, corinthians, rememb, enjooyyyyy, honeycomb, discernment, skillful, discouragement, omnipotent, 1cor, comforts, htgawm, heeeeeeyyyy, rewatches, beeeddd, 2cor, 2corinthians, 2thes, alyciamarie, contact_agency, godnat, uprightness, kavinsky, toooooo, familiytime, torontooooooooooooo, upright, twisting | BeautyPeachiii, mrsguendogan_, MamaPeachiii, AwesomenessTVde, LisaSch16, swaggiebiiebs, MarenloveShanti, lillipau, SAMINATORSELENA, DailyMandTNews | | |
| 15 | af1s, strut, aggiunge, buttercremetorte, rausbringst, scatti, altenheims, 600mm, zwangloses, kandern, autistinnen, joghurtkuchen, r3stocked, ellas, autocheckout, facelier, headioli, beistellponnys, kinderfreizeitkarte, mandarinenfüllung, nervigster, blutbucht, matzen, schockierendes, 2243, bahnwaggon, grenzbereich, dendeutsch, ausgesetzte, symoné, snkrs, eseln, 25x, supertele, mca, supersaftiger, 50zigern, kash, fascicoli, jaumann, pinch, acros, stallion, overpriced, fussballfan, 402, themself, carlotti, verwaiste, wrestlers | RTL_DSDS, Ben33711, HelmutOrosz, VanessaErmisch, Schokofestival, MalzJohannes, osmankrasnic63, loveM_selina, IlaydaBok, CrazySaraaaaaah | | |
| 14 | wandere, kopfausseelean, tweetapicturewithoutlooking, supportin, richertz, miyabi, mieseste, spielkinder, wembleyyyy, rumundehre, endfolgen, endfolgt, monumentales, kawai, twitterversary, teaaaaaaaa, kummerspeck, momjeans, errät, uhhhhhh, gnadenfrist, dranzuhängen, whaaaaaat, frustesse, heung, soe, summertimeball2019, howl, pixum, calum, cabbage, shawnmendesthetour, erlebnissen, signierstunde, hinterfrage, spieluhr, uhhhhh, voicemail, schnapsidee, wasserdampf, irwin, interacting, usp, bumst, filthy, temptation, bangin, wiege, sirens, häkchen | DouniaSlimani, Laura_Slimani, karin_slimani, Miriam2806, nadine_unicorn, _BabySteps_, Clarax3Sami, anh291000, fangirl_0305, mala_alisha | | |
| 14 | تضامنا, wundererzählungen, بالمغرب, تخرج, ومظهر, 2773, تركوا, تخفى, goldblitzendes, تحضر, تعقيبا, perennialgarden, تخريجة, goldgeprägt, goldgeprägten, goldgerahmten, rara, dekorativem, perserkönig, تا, perréal, تفكرت, ديال, wundervölker, königshauses, währendes, detailreiche, وجيز, خاطرو, beredte, حاط, حاضرة, وغادي, keilerei, حقوق, وراء, sagenstoff, جوج, informationsbroschüre, تنقل, salas, ثلاثين, normalausgabe, vorsatzblatt, goldschnitt, vitaler, تنتشر, pdfelement7release, يقاومون, altfranzösischen | twitt_erfolg_de, brandschutztip, gartenspritzeka, designermode_, outlet_tip, outlet__mode, SchuhXL, VDrabent, yamaghreb, hafid64 | | |

## Source Code of the Sampling Prototype

latest: https://github.com/FlxVctr/RADICES
equivalent to version used for this research: https://doi.org/10.6084/m9.figshare.8864777.v3